\PassOptionsToPackage{unicode}{hyperref}
\PassOptionsToPackage{hyphens}{url}

\documentclass{article}

\usepackage{arxiv}

\usepackage[utf8]{inputenc} %
\usepackage[T1]{fontenc}    %
\usepackage{hyperref}       %
\usepackage{url}            %
\usepackage{booktabs}       %
\usepackage{amsfonts}       %
\usepackage{nicefrac}       %
\usepackage{microtype}      %
\usepackage{cleveref}       %
\usepackage{lipsum}         %
\usepackage{graphicx}
\usepackage{doi}

\NewDocumentCommand\citeproctext{}{}
\NewDocumentCommand\citeproc{mm}{%
  \begingroup\def\citeproctext{#2}\cite{#1}\endgroup}
\makeatletter
 \let\@cite@ofmt\@firstofone
 \def\@biblabel#1{}
 \def\@cite#1#2{{#1\if@tempswa , #2\fi}}
\makeatother
\newlength{\cslhangindent}
\newlength{\csllabelwidth}
\newenvironment{CSLReferences}[2] %
 {\begin{list}{}{%
  \setlength{\itemindent}{0pt}
  \setlength{\leftmargin}{0pt}
  \setlength{\parsep}{0pt}
  \ifodd #1
   \setlength{\leftmargin}{\cslhangindent}
   \setlength{\itemindent}{-1\cslhangindent}
  \fi
  \setlength{\itemsep}{#2\baselineskip}}}
 {\end{list}}
\usepackage{calc}

\IfFileExists{bookmark.sty}{\usepackage{bookmark}}{\usepackage{hyperref}}
\IfFileExists{xurl.sty}{\usepackage{xurl}}{} %
\hypersetup{
  pdftitle={Deep Learning for Educational Data Science},
    pdfauthor={true, true},
  hidelinks,
  pdfcreator={LaTeX via pandoc}}

\title{Deep Learning for Educational Data Science}

\date{}

\newif\ifuniqueAffiliation
\uniqueAffiliationtrue

\ifuniqueAffiliation %
\author{\href{https://orcid.org/0000-0002-2972-485X}{\includegraphics[scale=0.06]{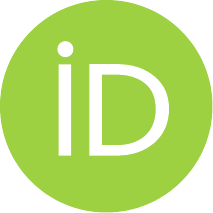}\hspace{1mm}Juan
D. Pinto} \\	University of Illinois
Urbana-Champaign\\	\texttt{jdpinto2@illinois.edu} \\
	\And
	\href{https://orcid.org/0000-0002-2738-3190}{\includegraphics[scale=0.06]{orcid.pdf}\hspace{1mm}Luc
Paquette} \\	University of Illinois
Urbana-Champaign\\	\texttt{lpaq@illinois.edu} \\}
\else
\usepackage{authblk}

\newbox{\orcid}\sbox{\orcid}{\includegraphics[scale=0.06]{orcid.pdf}}
\author[1]{%
	\href{https://orcid.org/0000-0000-0000-0000}{\usebox{\orcid}\hspace{1mm}David S.~Hippocampus\thanks{\texttt{hippo@cs.cranberry-lemon.edu}}}%
}
\author[1,2]{%
	\href{https://orcid.org/0000-0000-0000-0000}{\usebox{\orcid}\hspace{1mm}Elias D.~Striatum\thanks{\texttt{stariate@ee.mount-sheikh.edu}}}%
}
\affil[1]{Department of Computer Science, Cranberry-Lemon University, Pittsburgh, PA 15213}
\affil[2]{Department of Electrical Engineering, Mount-Sheikh University, Santa Narimana, Levand}
\fi

\renewcommand{\headeright}{(preprint)}
\renewcommand{\undertitle}{(preprint)}

\begin{document}
\maketitle

\begin{abstract}
With the ever-growing presence of deep artificial neural networks in
every facet of modern life, a growing body of researchers in educational
data science---a field consisting of various interrelated research
communities---have turned their attention~to leveraging these powerful
algorithms within the domain of~education. Use cases range from advanced
knowledge tracing models that can leverage open-ended student essays or
snippets of code to automatic affect~and~behavior detectors that can
identify when a student is frustrated or~aimlessly trying to solve
problems unproductively---and~much more. This chapter provides a brief
introduction to deep learning, describes some of its advantages and
limitations, presents a survey of its many uses in education, and
discusses how it may further come to shape the field of educational data
science.
\end{abstract}

\section{Introduction}\label{introduction}

As artificial intelligence (AI) continues to penetrate ever deeper into
modern life, one particular family of machine learning
algorithms---namely, deep neural networks---have come to be seen as the
solution to many of the challenges that have stumped more classical
algorithms in the past. Modeled loosely on the structure of
\emph{biological} neural networks, \emph{artificial} neural networks
consist of chains of simple mathematical transformations that can model
complex non-linear decision boundaries in large problem spaces. In
particular, \emph{deep} neural networks---artificial neural networks
that consist of multiple layers of transformations---allow for
sufficient complexity to tackle tasks in a wide variety of fields. These
models are collectively and more colloquially referred to as deep
learning.

A growing body of education researchers are now also turning their
attention~to leveraging the power of deep learning algorithms for the
tasks of improving and understanding human learning. Researchers in
educational data science, a field consisting of various interrelated
research communities such as Educational Data Mining~(EDM), Learning
Analytics (LA), and AI in Education (AIED), have been involved in this
endeavor. Given the histories of these communities and their goals, the
contexts and rationales behind the use of deep learning in education are
highly varied. It has been used in K--12
(\citeproc{ref-somExplainableStudentGroup2021}{Som et al., 2021};
\citeproc{ref-southwellChallengesFeasibilityAutomatic2022}{Southwell et
al., 2022}), higher education
(\citeproc{ref-rashidLecturerPerformanceSystem2016}{Rashid \& Ahmad,
2016}; \citeproc{ref-zhangUndergraduateGradePrediction2021}{Y. Zhang et
al., 2021}), and non-traditional online education
(\citeproc{ref-chenApplyingRecentInnovations2020}{Chen \& Pardos, 2020};
\citeproc{ref-fengUnderstandingDropoutsMOOCs2019}{Feng et al., 2019}).
Within these contexts, it has been used to tackle different tasks,
including predicting students' future actions
(\citeproc{ref-fengUnderstandingDropoutsMOOCs2019}{Feng et al., 2019};
\citeproc{ref-qiuBetterGradePrediction2022}{Qiu et al., 2022}),
knowledge tracing (\citeproc{ref-piechDeepKnowledgeTracing2015}{Piech et
al., 2015}; \citeproc{ref-puSelfattentionKnowledgeTracing2022}{Pu \&
Becker, 2022}), automated assessment
(\citeproc{ref-tanAutomaticShortAnswer2020}{H. Tan et al., 2020};
\citeproc{ref-taySkipFlowIncorporatingNeural2018}{Tay et al., 2018}),
affect detection
(\citeproc{ref-botelhoStudyingAffectDynamics2018}{Botelho et al., 2018};
\citeproc{ref-lanAccurateInterpretableSensorfree2020}{Lan et al.,
2020}), and recommendation systems
(\citeproc{ref-jiangGoalbasedCourseRecommendation2019}{W. Jiang et al.,
2019}; \citeproc{ref-shenAutomaticRecommendationTechnology2016}{Shen et
al., 2016}), to name a few use cases. Yet, despite the recent upsurge in
educational publications using deep learning, many challenges remain
that make it difficult for existing learning platforms to adopt these
models outside of research laboratories.

In this chapter, we first provide a brief introduction to machine
learning and place deep learning in its greater context. We then
highlight the advantages and limitations of using deep learning in
education. We follow this with a survey of how this novel technology has
been applied, with the section organized around specific use cases.
Finally, we discuss what the future of deep learning in education may
hold, including the principal challenges that must be overcome to
advance how these algorithms impact learners at scale.

\section{Deep learning in context}\label{deep-learning-in-context}

To understand how deep learning is being used for educational data
science, as well as its future potential, it is helpful to have a
cursory familiarity with machine learning more broadly. Machine learning
models are algorithms that can predict or estimate the value of a target
variable of interest (predicted variable, or output) given data related
to the problem (predictor variables, or inputs). Through the process of
`training' on this data (the \emph{training dataset}), a model begins to
identify the patterns of inputs that lead to specific outputs. Once
trained, the model has `learned' these patterns and can be used to make
predictions on inputs it has never seen before. The model's accuracy is
then measured through a series of metrics that compare its predictions
on this \emph{testing dataset} to the true values of the target variable
being predicted.

Deep learning is only one subset of the larger family of machine
learning models, which include many other algorithms, such as
\emph{random forest}, \emph{K-nearest neighbors}, \emph{support vector
machines}, and \emph{linear regression}. Deep learning is itself a group
composed of different model architectures which have been developed and
refined over the years to suit different purposes. A deep learning
model's architecture refers to the structures that define how artificial
neurons are connected to each other and the different parameters that
are used to model such connections.

Many different neural architectures have been used in education. Fully
connected (or dense) neural networks connect each input node to each
output node for each layer. These are simpler architectures that are
most often used as part of a larger model but are rarely used as
stand-alone models. Convolutional neural networks (CNNs) include at
least one convolutional layer, in which a sliding window shifts across
the input vector or matrix and a dot product is calculated. These are
most commonly used for image inputs, but they can also be used with
sequential data. Recurrent neural networks (RNNs) have at least a single
recurrent layer, in which a looping function processes a sequence of
inputs---one item at a time---while carrying information from previous
items to future items in the sequence. Two common variations of the RNN
architecture are the long short-term memory network (LSTM) and the gated
recurrent unit (GRU), which include slight memory modifications to
improve performance. Transformers, another common neural network
architecture, convert inputs into vectors called encodings and then
decode them into different types of outputs depending on the task. These
are used with sequential data (most commonly natural language text) and
rely on a neural attention mechanism to identify the relevance of
neighboring data when encoding a particular input
(\citeproc{ref-vaswaniAttentionAllYou2017}{Vaswani et al., 2017}).
Popular generative models such as GPT-4 and DALL-E are based on this
transformer architecture. Graph neural networks (GNNs) take graphs as
inputs (data organized into nodes connected by edges) and create a
separate network for each node (and, optionally, each edge) in the
graph. Each sub-network makes its own calculations based on the
neighbors of its pertaining node. Information can be shared between
parts of the graph and pooled together for final prediction. For a more
in-depth survey of the most common deep learning architectures, see Alom
et al. (\citeproc{ref-alomStateoftheartSurveyDeep2019}{2019}) and
Alzubaidi et al. (\citeproc{ref-alzubaidiReviewDeepLearning2021}{2021}).

\section{Why deep learning in
education}\label{why-deep-learning-in-education}

\subsection{Advantages}\label{advantages}

The widespread adoption of deep learning suggests that these models hold
certain advantages over more traditional data-driven approaches to
prediction, clustering, and analysis tasks. Overall, deep learning
provides greater flexibility compared to other machine learning
approaches. This flexibility manifests itself in the increased
predictive power that is made possible by having such a large solution
space to work with. It also allows for models that require less direct
human intervention during training, that can use more varied types of
input data, and for which training can easily be paused and resumed at a
later time when more data is available.

\subsubsection{Increased predictive
accuracy}\label{increased-predictive-accuracy}

Among the many studies in education that have compared deep learning
models to more traditional algorithms, most of them have reported an
overall increase in accuracy. The accuracy of deep neural networks can
be largely attributed to the large number of learnable parameters that
these models have---from tens of thousands to millions or even billions.
Improvements in hardware, along with the large amounts of data that are
now regularly collected through digital platforms, have made it possible
to train models that can fit very complex functions. These models can
also project data to high-dimensional spaces, which can help with
identifying patterns.

Data-driven approaches in education (including both deep learning and
more traditional algorithms) are only as good as their ability to
accurately model the real world. For example, in knowledge tracing---the
task of predicting how students will perform in future problems based on
an estimation of what they have learned in previous
problems---traditional algorithms usually treat individual parameters
and skills as independent from each other
(\citeproc{ref-corbettKnowledgeTracingModeling1995}{Corbett \& Anderson,
1995}). This naive approach is easier to implement, interpret, and
optimize, but it is a simplified assumption of reality that deep
learning approaches don't necessarily make. From this perspective,
predictive accuracy serves as a proxy for how well the learning process
has been simulated. This makes increased accuracy an important goal to
pursue, and something with which deep learning can help. Of course,
beyond the architecture and details of the model itself, other aspects
can have a big impact on accuracy, including the amount and quality of
available data and the way features are engineered.

\subsubsection{Automatic feature
engineering}\label{automatic-feature-engineering}

Feature engineering refers to the act of extracting predictor variables
from raw data that are useful for a prediction task. Despite tools that
can automatically create large sets of features through simple
transformations (\citeproc{ref-kanterDeepFeatureSynthesis2015}{Kanter \&
Veeramachaneni, 2015}), feature engineering continues to be a largely
human-driven and inexact `art' more than a science. It often requires
extensive experience and domain expertise, and it can also be a very
time-consuming part of the process of developing data-driven models.

As opposed to traditional machine learning techniques that make
predictions directly from a set of pre-engineered features, deep
learning techniques conduct their own representation learning through
multi-step transformations that can create very complex features. These
models learn which transformations to their inputs lead to accurate
predictions. Much of the increased predictive accuracy of deep neural
networks comes from this ability to learn complex high-level features
from raw, low-level data. This ability bypasses the need for
resource-intensive feature engineering.

\subsubsection{Flexible inputs}\label{flexible-inputs}

Related to the idea of automatic feature engineering is deep learning's
ability to use a wider variation of inputs than traditional algorithms.
This helps explain why deep learning has gained such a strong reputation
for computer vision and speech recognition tasks
(\citeproc{ref-alamSurveyDeepNeural2020}{Alam et al., 2020})---video and
sound data are difficult to convert to useful tabular features, but they
can be used directly as raw inputs in a neural network.

In educational data science, input flexibility has implications for
multimodal learning analytics, which by definition involves data of
different modalities. It also makes it possible to directly analyze
students' responses to open-ended questions, such as for automated essay
and short answer scoring
(\citeproc{ref-dascaluReaderBenchEnvironmentAnalyzing2013}{Dascalu et
al., 2013}; \citeproc{ref-tanAutomaticShortAnswer2020}{H. Tan et al.,
2020}; \citeproc{ref-taySkipFlowIncorporatingNeural2018}{Tay et al.,
2018}), the scoring of open responses in math education
(\citeproc{ref-baralImprovingAutomatedScoring2021}{Baral et al., 2021};
\citeproc{ref-ericksonAutomatedGradingStudent2020}{Erickson et al.,
2020}), and the direct analysis of student code in computer science
education (\citeproc{ref-shiMoreLessExploring2021}{Shi, Mao, et al.,
2021}; \citeproc{ref-shiSemiautomaticMisconceptionDiscovery2021}{Shi,
Shah, et al., 2021}). This same flexibility can also apply to outputs,
making it possible to predict sequential outputs of different
kinds---useful for providing writing suggestions, code hints, and
automated targeted feedback.

\subsubsection{Continuous model
training}\label{continuous-model-training}

Traditional machine learning models are trained on an entire dataset and
then deployed for use. If new data is made available, the model must
typically be retrained from scratch to get the best results. Some
traditional machine learning algorithms, such as tree-based models, do
make it possible to introduce new data---but only in a limited fashion
that does not alter the pre-existing learned parameters. Deep neural
networks, on the other hand, do not have this limitation. Because they
are trained over multiple epochs (rounds of training), introducing new
data simply means picking up where training left off with additional
epochs, using either the new data exclusively or a combined dataset of
old and new data.

The ability to update models is especially important when deployed in
real-world scenarios that may require periodic adjustment to combat
concept drift (the changes that input-output patterns undergo as time
progresses) or simply to improve accuracy. For example, the
effectiveness of an intelligent tutoring system (ITS) may be improved if
its underlying algorithm can be updated after it has been in use for
some time. New data on student interactions can provide an accuracy
boost, or they can be used to capture changes in how students interact
with the learning platform following changes in the platform's design.

\subsubsection{Transfer learning}\label{transfer-learning}

Taking continuous model training a step further, transfer learning is a
machine learning approach in which a model that has been trained in one
domain and for a specific task is then reused in either a different
domain or for a different task
(\citeproc{ref-zhuangComprehensiveSurveyTransfer2021}{Zhuang et al.,
2021}). The goal is to leverage the knowledge learned from one domain to
improve the performance of the model on the other. Often, there is
abundant labeled data for the source domain and only limited labeled
data for the target domain.

While transfer learning is possible with traditional algorithms, the
ability to continue training a deep neural network makes more powerful
transfer learning possible. A trained model can be partly frozen (only a
subset of its parameters is retrained), it can be expanded with
additional layers that can completely reshape its architecture, or it
can be used largely as is with only minor modifications, based on the
needs of the target domain and task. The recurring theme of flexibility
applies here as well. In education, transfer learning may one day lead
to more generalizable models that can be used across different
platforms, with different populations, or for different courses.
Transfer learning can also be helpful in situations where labeled data
is limited---not a rare occurrence in education.

\subsection{Limitations}\label{limitations}

As we will see in the next section, deep learning is being used more and
more for educational data science, but it has yet to significantly shape
our understanding of the learning process or to affect large numbers of
students and instructors in real-world settings. Instead, it has been
the subject of rigorous study in university laboratories. While this may
partly be attributed to its novelty, it is also likely that the
limitations inherent in the technology have played a role. Here we
discuss some of the biggest limitations of deep learning and how these
may apply in the domain of education.

\subsubsection{Diminished
interpretability}\label{diminished-interpretability}

When compared with simpler, more traditional models, the principal
limitation of deep neural networks is their inherent lack of
transparency. Due to their complexity, it is often impossible to
determine why a model is making specific decisions. This ability to peel
back the curtain on a model's inner workings is often referred to as
interpretability
(\citeproc{ref-cohauszWhenProbabilitiesAre2022}{Cohausz, 2022}). In
fields such as education, where implications can be serious and many
stakeholders are involved (students, parents, teachers, administrators,
politicians), interpretability can be of utmost importance, carrying
implications for trustworthiness, accountability, and trust. This means
that where performance is similar between models, or where
interpretability is more important than an increase in accuracy, it is
often preferable to go with more traditional, interpretable approaches.

At a more general level, there remain fundamental questions about how
exactly some deep learning algorithms are able to do what they do. It
has been noted that \emph{untrained} neural networks can surprisingly
perform almost as well as trained ones for specific use cases
(\citeproc{ref-botelhoDeepLearningDeep2022}{Botelho et al., 2022};
\citeproc{ref-dingWhyDeepKnowledge2019}{Ding \& Larson, 2019}). This
serves as a startling reminder of the black-box nature of these models,
and it is easy to see why caution must be exerted when making claims
about their understanding of the world. On the other hand, the fact that
some can identify patterns so effectively, even when untrained, hints at
some possible untapped potential. Moreover, these models can still be
useful in cases where predictive accuracy trumps interpretability
(\citeproc{ref-botelhoDeepLearningDeep2022}{Botelho et al., 2022}) and
in the classic EDM task of `discovery with models'
(\citeproc{ref-bakerStateEducationalData2009}{Baker \& Yacef, 2009}).

\subsubsection{Heightened model
complexity}\label{heightened-model-complexity}

As stated, the reason for diminished interpretability in deep neural
networks lies in their complexity. This complexity can come both in
terms of model architecture and number of parameters. Even `simple'
neural networks with only one or two hidden layers can consist of
thousands of learnable parameters, with more complex ones having orders
of magnitude more. Aside from the issue of interpretability that this
creates, it can also make designing, training, and using these models
more resource intensive. As Y. Jiang et al.
(\citeproc{ref-jiangExpertFeatureengineeringVs2018}{2018}) have pointed
out, deep learning models are often used for the time- and
effort-savings promised by their automatic feature engineering. However,
the act of refining complex architectures and effectively selecting
proper hyperparameters (unlearnable, human-set parameters that determine
how a model functions) may negate this advantage, especially when
adapting models to new domains or tasks. It may be that further research
reveals consistent best practices and ideal models for varying
situations, but for now this remains a largely open-ended task in
itself.

\subsubsection{Need for large datasets}\label{need-for-large-datasets}

The heightened model complexity of deep neural networks also leads to an
increased requirement for large datasets. It is generally accepted that
the more parameters a model has, the more data is needed to adequately
train it. This is a major reason why deep learning has only taken off
after the internet made big data more easily available, despite the
principal theories behind the technology having been known about for
many decades (\citeproc{ref-tappertWhoFatherDeep2019}{Tappert, 2019}).

In educational data science, this makes deep learning impractical for
tasks and domains where not enough data exists. Data scarcity is
particularly common in use cases that require labeled data (i.e.~data
that has been tagged with meaningful information) due to the resources
required to label large amounts of data. Studies have found that deep
learning models are more likely to overfit when datasets are smaller
(\citeproc{ref-gervetWhenDeepLearning2020}{Gervet et al., 2020}).
Besides the obvious (but not always possible) solution of simply
collecting more data, there exist ways to make deep learning useful when
data is limited---such as transfer learning, semi-supervised learning
(\citeproc{ref-livierisPredictingSecondarySchool2019}{Livieris et al.,
2019}; \citeproc{ref-shiMoreLessExploring2021}{Shi, Mao, et al., 2021}),
or data augmentation (\citeproc{ref-caderPotentialUseDeep2020}{Cader,
2020}). Still, researchers rely on large datasets when available, while
keeping in mind that high-quality data can be more effective than big,
noisy data (\citeproc{ref-yudelsonBetterDataBeat2014}{Yudelson et al.,
2014}).

\subsubsection{Expanded risk from expanded
scale}\label{expanded-risk-from-expanded-scale}

Lastly, many of the same issues faced by traditional machine learning
algorithms also apply to deep learning, simply at a bigger scale.
Because of the bigger data requirements, privacy and security breaches
have the potential to affect more people, more quickly, and more deeply.
This can be especially damaging when sensitive data is involved, such as
video or speech recordings, which are more likely to be used with deep
learning models due to their ability to process this raw data. The
increase in predictive ability for tasks such as tracking,
identification, or behavior modeling also increases the potential that
such tools will be maliciously used for surveillance. There are
documented cases of algorithmic bias and unfairness in education, in
which biased data disproportionally and negatively affects students from
historically disadvantaged populations
(\citeproc{ref-bakerAlgorithmicBiasEducation2022}{Baker \& Hawn, 2022};
\citeproc{ref-kizilcecAlgorithmicFairnessEducation2022}{Kizilcec \& Lee,
2022}). It is clear that there is still much we don't know about the
real-life implications of these biases, and the black-box nature of deep
neural networks further obscures our understanding.

While none of these risks are exclusive to deep learning, their
increased potential for damage requires that ongoing work be done to
find ways to mitigate their reach. Until researchers find ways to
appropriately do so, the expanded risks and diminished interpretability
of deep learning suggest that it may be unwise---and in some cases even
unethical---to use them for high-stakes educational tasks. Luckily, this
is an area of research that is receiving increasing levels of attention
and interest, both inside and outside of education.

\section{A survey of deep learning in
education}\label{a-survey-of-deep-learning-in-education}

To better understand the various ways that deep learning has been used
in education, this section organizes the literature into two separate
categories: \emph{direct} and \emph{indirect} uses. By direct uses, we
mean contexts in which a specific problem or task is tackled for which
deep learning plays a central role. For example, in knowledge tracing,
the problem is measuring what a student knows or what they may struggle
with based on their past actions. In affect detection, the problem is
measuring the various emotions (and consequently the level of engagement
and motivation) that a student may be experiencing. In both of these
cases, deep learning can be the mechanism itself by which the problem is
tackled.

Indirect uses, on the other hand, make use of deep learning only as a
steppingstone towards a separate main goal. For example, useful features
can be extracted by leveraging the automatic feature engineering that
deep learning models conduct, which can then be used as the inputs for a
separate algorithm to conduct the prediction or analysis desired
(\citeproc{ref-ericksonAutomatedGradingStudent2020}{Erickson et al.,
2020}). Similarly, deep learning can be used to transcribe student
speech, which can then be analyzed using a variety of different methods
(\citeproc{ref-pughSayWhatAutomatic2021}{Pugh et al., 2021}). Indirect
uses are typically tasks that have been very successfully performed by
deep learning in other domains, and which are now being leveraged for
educational purposes. In many of these cases, ``off-the-shelf'' models
already exist.

The difference between these two categories may not be immediately
clear, but it can be helpful for understanding the efforts required for
implementation. Note that the distinction is mainly about the immediate
purpose for which the deep learning model was designed. In one case, a
deep learning model specifically and directly tackles an educational
task (e.g.~assess student writing), whereas in the other, the model
predicts something other than the project's ultimate educational goal
(e.g.~transcribe student speech). The outputs of the indirect models
serve as inputs to a separate analysis in order to arrive at the
phenomenon of interest. In both direct and indirect uses of deep
learning, a deep neural network is used somewhere along the pipeline,
but the specific architecture or design of the network can vary
drastically depending on the inputs and desired outputs.

\subsection{Direct uses of deep learning in
education}\label{direct-uses-of-deep-learning-in-education}

\subsubsection{Predicting future
actions}\label{predicting-future-actions}

Models for future prediction are those that use data gathered up to a
certain point in time as input and make a prediction about some event
that will take place after said time. The most common use cases for
these models have been to predict student grades and student dropout.
Multiple studies have used past grades as the primary or exclusive
predictor of future grades, including some using the LSTM architecture
with a neural attention mechanism
(\citeproc{ref-huAcademicPerformanceEstimation2019}{Q. Hu \& Rangwala,
2019}; \citeproc{ref-qiuBetterGradePrediction2022}{Qiu et al., 2022}),
as well as more traditional fully connected networks
(\citeproc{ref-livierisPredictingStudentsPerformance2012}{Livieris et
al., 2012};
\citeproc{ref-livierisPredictingSecondarySchool2019}{Livieris et al.,
2019}). Y. Zhang et al.
(\citeproc{ref-zhangUndergraduateGradePrediction2021}{2021})
interestingly used a CNN with a neural attention mechanism, with course
descriptions and demographic features as input data.

Student dropout prediction has been thoroughly studied in massive open
online courses (MOOCs) using clickstream data (often called log data)
gathered from these digital platforms. Deep learning architectures for
this purpose have included fully connected networks
(\citeproc{ref-whitehillDelvingDeeperMOOC2017}{Whitehill et al., 2017}),
CNNs with neural attention
(\citeproc{ref-fengUnderstandingDropoutsMOOCs2019}{Feng et al., 2019}),
and LSTMs (\citeproc{ref-feiTemporalModelsPredicting2015}{Fei \& Yeung,
2015}). As an example of deep learning\textquotesingle s ability to make
use of highly varied data, Cohausz
(\citeproc{ref-cohauszWhenProbabilitiesAre2022}{2022}) used student
behavioral features, classes taken, past grades, and demographic
information to predict student dropout in a mandatory computer science
course. Similarly, Swamy et al.
(\citeproc{ref-swamyEvaluatingExplainersBlackbox2022}{2022}) used
measures of engagement, student participation, and study habits to
predict students\textquotesingle{} probability of passing a MOOC.

Aside from these predictions of course success, deep learning has also
been used to predict future course enrollment
(\citeproc{ref-doleckPredictiveAnalyticsEducation2020}{Doleck et al.,
2020}), college success for admissions purposes
(\citeproc{ref-oladokunPredictingStudentsAcademic2008}{Oladokun, 2008}),
students' next action on an online course
(\citeproc{ref-chenApplyingRecentInnovations2020}{Chen \& Pardos, 2020};
\citeproc{ref-tangDeepNeuralNetworks2016}{Tang et al., 2016}), program
graduation (\citeproc{ref-kimGritNetStudentPerformance2018}{Kim et al.,
2018}), and learning gains (\citeproc{ref-linComparisonsBKTRNN2017}{Lin
\& Chi, 2017}). Yeung \& Yeung
(\citeproc{ref-yeungIncorporatingFeaturesLearned2019}{2019})
interestingly used the last hidden state of an LSTM along with student
profile features to predict whether students would choose a STEM or
non-STEM job after college. Jarbou et al.
(\citeproc{ref-jarbouDeepLearningbasedSchool2022}{2022}) were able to
predict short- and long-term school absenteeism among students with
autism spectrum disorder---a population that is disproportionally
affected by school absenteeism and for whom missing class can be
especially detrimental
(\citeproc{ref-gottfriedChronicAbsenteeismIts2014}{Gottfried, 2014};
\citeproc{ref-munkhaugenIndividualCharacteristicsStudents2019}{Munkhaugen
et al., 2019}).

\subsubsection{Knowledge tracing}\label{knowledge-tracing}

Knowledge tracing (KT), or knowledge inference, refers to the process of
estimating what a learner knows at any given time and how that knowledge
changes over time. Because knowledge is not a directly observable trait,
the process is also known as latent knowledge estimation. KT ultimately
serves a different purpose than predicting future actions. However,
because of the latent nature of student knowledge, predictive accuracy
on the performance of subsequent tasks is typically the defining
characteristic by which the success of a KT model is measured.

Historically, KT has been dominated by approaches such as Bayesian
Knowledge Tracing (BKT;
\citeproc{ref-corbettKnowledgeTracingModeling1995}{Corbett \& Anderson,
1995}) and its many variants
(\citeproc{ref-bakerMoreAccurateStudent2008}{Baker et al., 2008};
\citeproc{ref-gonzalez-brenesGeneralFeaturesKnowledge2014}{Gonzalez-Brenes
et al., 2014};
\citeproc{ref-khajahIntegratingLatentfactorKnowledgetracing2014}{M. M.
Khajah et al., 2014};
\citeproc{ref-pardosDeterminingSignificanceItem2009}{Pardos \&
Heffernan, 2009},
\citeproc{ref-pardosModelingIndividualizationBayesian2010}{2010},
\citeproc{ref-pardosKTIDEMIntroducingItem2011}{2011};
\citeproc{ref-yudelsonIndividualizedBayesianKnowledge2013}{Yudelson et
al., 2013}). BKT estimates learners' knowledge using a hidden Markov
model, dynamically updating latent variables based on ongoing student
performance. Other popular techniques include performance factors
analysis (PFA;
\citeproc{ref-pavlikjrPerformanceFactorsAnalysis2009}{Pavlik Jr et al.,
2009}) and item response theory (IRT;
\citeproc{ref-raschProbabilisticModelsIntelligence1993}{Rasch, 1993}).
While these approaches are still commonly used, much of the research
focus on KT has shifted in recent years to ever-more-complex deep
learning models.

In the first paper that utilized deep learning for knowledge tracing
research, Piech et al.
(\citeproc{ref-piechDeepKnowledgeTracing2015}{2015}) described two
models, which they named deep knowledge tracing (DKT)---one was a
standard RNN while the other used an LSTM. Both models relied
exclusively on prior student performance as inputs---an approach that
follows the example of the classical BKT approach. Their positive
results showed the potential of deep learning knowledge tracing (DLKT)
models, even after accounting for issues in their methods and reporting
(\citeproc{ref-khajahHowDeepKnowledge2016}{M. Khajah et al., 2016};
\citeproc{ref-xiongGoingDeeperDeep2016}{Xiong et al., 2016}), sparking a
race toward ever-more-elaborate uses of deep learning for KT.

Since then, the many DLKT models that have been developed thus far can
be categorized into five general groups based on their model
architectures and intended inputs
(\citeproc{ref-liuSurveyKnowledgeTracing2023}{Q. Liu et al., 2023};
\citeproc{ref-sarsaEmpiricalEvaluationDeep2022a}{Sarsa et al., 2022}).
First, RNN-based models are those most similar to the original DKT by
Piech et al. (\citeproc{ref-piechDeepKnowledgeTracing2015}{2015}) in
that they consist of at least one recurrent layer (eg.
\citeproc{ref-delianidiStudentPerformancePrediction2021}{Delianidi et
al., 2021}; \citeproc{ref-xiongGoingDeeperDeep2016}{Xiong et al., 2016})
and variations such as LSTM (eg.
\citeproc{ref-maoDeepLearningVs2018}{Mao et al., 2018};
\citeproc{ref-penmetsaInvestigateEffectivenessCode2021}{Penmetsa et al.,
2021}; \citeproc{ref-shiCodeDKTCodebasedKnowledge2022}{Shi et al.,
2022}; \citeproc{ref-tatoDeepKnowledgeTracing2022}{Tato \& Nkambou,
2022}; \citeproc{ref-wangLearningRepresentStudent2017}{L. Wang et al.,
2017}). Second, memory-based models attempt to address the lack of
explicit knowledge component (KC) tracking by introducing a memory
module to the architecture (eg.
\citeproc{ref-abdelrahmanKnowledgeTracingSequential2019}{Abdelrahman \&
Wang, 2019}; \citeproc{ref-aiConceptawareDeepKnowledge2019}{Ai et al.,
2019}; \citeproc{ref-karumbaiahUsingNeuralNetworkbased2022}{Karumbaiah
et al., 2022}; \citeproc{ref-zhangDynamicKeyvalueMemory2017}{Zhang et
al., 2017}). This allows them to align more closely with the intended
goal of KT rather than simply relying on predictive ability. Third,
exercise-based models include additional information about each problem
as input, which is gathered from the text of the problem using natural
language processing (NLP) techniques (eg.
\citeproc{ref-liuEKTExerciseawareKnowledge2021}{Q. Liu et al., 2021};
\citeproc{ref-liuImprovingKnowledgeTracing2020}{Y. Liu et al., 2020};
\citeproc{ref-suExerciseenhancedSequentialModeling2018}{Su et al.,
2018}). Fourth, attention-based models make use of neural attention to
weight the relevance of surrounding actions (eg.
\citeproc{ref-choiAppropriateQueryKey2020}{Choi et al., 2020};
\citeproc{ref-ghoshContextawareAttentiveKnowledge2020}{Ghosh et al.,
2020}; \citeproc{ref-pandeySelfattentiveModelKnowledge2019}{Pandey \&
Karypis, 2019};
\citeproc{ref-pandeyRKTRelationawareSelfattention2020}{Pandey \&
Srivastava, 2020}; \citeproc{ref-puSelfattentionKnowledgeTracing2022}{Pu
\& Becker, 2022}; \citeproc{ref-shinSAINTIntegratingTemporal2021}{Shin
et al., 2021}). Exercise-based models also make use of neural attention,
as does the memory-based model described by Abdelrahman \& Wang
(\citeproc{ref-abdelrahmanKnowledgeTracingSequential2019}{2019}). Fifth,
graph-based models use graph-based data as inputs (eg.
\citeproc{ref-longAutomaticalGraphbasedKnowledge2022}{Long et al.,
2022}; \citeproc{ref-nakagawaGraphbasedKnowledgeTracing2019}{Nakagawa et
al., 2019}; \citeproc{ref-songJKTJointGraph2021}{Song et al., 2021};
\citeproc{ref-tongStructurebasedKnowledgeTracing2020}{Tong et al.,
2020}). Graphs can leverage the interconnected nature of learning from
various angles, such as similarities between KCs or hierarchical course
structures.

\subsubsection{Automated assessment}\label{automated-assessment}

Automated assessment has been conducted in many forms, most recently
with deep learning models. Automated essay scoring and short answer
scoring have been attempted with LSTMs, fully connected networks,
bidirectional transformers, CNNs, memory networks, and GNNs {[}Taghipour
\& Ng (\citeproc{ref-taghipourNeuralApproachAutomated2016}{2016});
Dascalu et al.
(\citeproc{ref-dascaluReaderBenchMultilingualFramework2017}{2017}); Zhao
et al. (\citeproc{ref-zhaoMemoryaugmentedNeuralModel2017}{2017});
Riordan et al.
(\citeproc{ref-riordanInvestigatingNeuralArchitectures2017}{2017}); Tay
et al. (\citeproc{ref-taySkipFlowIncorporatingNeural2018}{2018}); H. Tan
et al. (\citeproc{ref-tanAutomaticShortAnswer2020}{2020})).
ReaderBench---an open-source multilingual text analysis tool
(\citeproc{ref-dascaluReaderBenchEnvironmentAnalyzing2013}{Dascalu et
al., 2013})---uses deep learning to automatically evaluate student
summaries of texts based on how well they covered the main idea of a
reading (\citeproc{ref-botarleanuAutomatedSummaryScoring2021}{Botarleanu
et al., 2021}).

Another use case involves peer evaluations, which often play an
important role in many large-scale courses such as MOOCs. However, it
can be difficult to evaluate the rigor of peer reviews at scale. To help
address this, Xiao et al.
(\citeproc{ref-xiaoDetectingProblemStatements2020}{2020}) trained
various deep and traditional models to identify whether reviewers'
responses to rubric items explicitly suggest ways for the author to
improve their work---an important aspect of high quality peer reviews.
Taking a different approach, Namanloo et al.
(\citeproc{ref-namanlooImprovingPeerAssessment2022}{2022}) used a GNN to
predict the proper score that a student's project should receive based
on peer evaluations, essentially acting as a peer assessment aggregation
mechanism that emulates expert scorers.

Within the realm of collaborative learning, T. Hu et al.
(\citeproc{ref-huAssessingStudentContributions2020}{2020}) used a CNN to
automatically score student edits on a collaborative Wiki assignment.
Meanwhile, Som et al.
(\citeproc{ref-somExplainableStudentGroup2021}{2021}) used a temporal
residual network (\citeproc{ref-wangTimeSeriesClassification2017}{Z.
Wang et al., 2017})---a specific CNN architecture that leverages
residual connections---to assess overall group collaboration quality
using individual students' changing roles within the group as input.

Automated assessment has also been tackled in the domain of computer
science education. Shi, Shah, et al.
(\citeproc{ref-shiSemiautomaticMisconceptionDiscovery2021}{2021})
trained a code2vec model---a specific architecture using neural
attention and designed to create meaningful embeddings of programming
code (\citeproc{ref-alonCode2vecLearningDistributed2019}{Alon et al.,
2019})---to predict the accuracy of student code submissions to
programming problems. Shi, Mao, et al.
(\citeproc{ref-shiMoreLessExploring2021}{2021}) compared the efficacy of
code2vec with its competitor Abstract Syntax Tree based Neural Network
(ASTNN; \citeproc{ref-zhangNovelNeuralSource2019}{Zhang et al., 2019})
at automatically detecting and classifying bugs in students' code.

\subsubsection{Affect detection}\label{affect-detection}

Student affect refers to the various emotions that students experience
as they undergo the learning process, such as delight, engagement,
surprise, confusion, frustration, and boredom
(\citeproc{ref-dmelloDynamicsAffectiveStates2012}{D'Mello \& Graesser,
2012}). Affect has been correlated with academic performance and
achievement (\citeproc{ref-craigAffectLearningExploratory2004}{Craig et
al., 2004}; \citeproc{ref-pardosAffectiveStatesState2014}{Pardos et al.,
2014}; \citeproc{ref-rodrigoAffectiveBehavioralPredictors2009}{Rodrigo
et al., 2009}) and has even been found to be predictive of later college
attendance (\citeproc{ref-sanpedroPredictingCollegeEnrollment2013}{San
Pedro et al., 2013}). Affect has often been detected using a variety of
physical and physiological sensors. Deep learning has been successfully
used to detect student affect using only log data---a practice known as
sensor-free affect detection
(\citeproc{ref-botelhoImprovingSensorfreeAffect2017}{Botelho et al.,
2017}; \citeproc{ref-botelhoStudyingAffectDynamics2018}{Botelho et al.,
2018}; \citeproc{ref-lanAccurateInterpretableSensorfree2020}{Lan et al.,
2020}). However, this is also possible using simple models such as
logistic regression, and it has been demonstrated that each approach
brings its own advantages
(\citeproc{ref-jiangExpertFeatureengineeringVs2018}{Y. Jiang et al.,
2018}).

\subsubsection{Recommendation systems}\label{recommendation-systems}

Recommendation systems are one of the most pervasive use-cases for AI in
modern daily life, and research has been devoted to bringing them to
education. In the deep learning space, Shen et al.
(\citeproc{ref-shenAutomaticRecommendationTechnology2016}{2016}) trained
a CNN model that can recommend learning resources to students in online
courses. The model can automatically identify the latent factors of
learning resources by analyzing the textual information they contain.

In another study, W. Jiang et al.
(\citeproc{ref-jiangGoalbasedCourseRecommendation2019}{2019}) devised an
algorithm that recommends to students certain prerequisite courses to
take in order to succeed in a future course they wish to take, based on
their past grades and the other courses in which they will enroll.
Reinforcement learning has also been used, such as by Ai et al.
(\citeproc{ref-aiConceptawareDeepKnowledge2019}{2019}) who trained a
model that can recommend the next exercises for students to attempt
based on their estimated latent knowledge. To do this, they trained the
model on a student simulation that used a KT model to imitate the
actions of a real student.

\subsubsection{Behavior detection}\label{behavior-detection}

Deep learning models have been used to automatically detect different
student behaviors, such as wheel spinning
(\citeproc{ref-botelhoDevelopingEarlyDetectors2019}{Botelho et al.,
2019}; \citeproc{ref-matsudaHowQuicklyCan2016}{Matsuda et al.,
2016})---a form of unproductive persistence
(\citeproc{ref-beckWheelspinningStudentsWho2013}{Beck \& Gong, 2013}).
Pinto et al. (\citeproc{ref-pintoInterpretableNeuralNetworks2023}{2023})
created an interpretable CNN-based detector of gaming the system
behavior, which has been defined as `attempting to succeed in an
interactive learning environment by exploiting properties of the system
rather than by learning the material'
(\citeproc{ref-bakerLabelingStudentBehavior2008}{Baker \& de Carvalho,
2008}). Deep learning has also been used to detect self-regulated
learning (SRL) behaviors in a collaborative context
(\citeproc{ref-nguyenExploringSociallyShared2022}{A. Nguyen et al.,
2022}). For detecting off-task behavior, Y. Jiang et al.
(\citeproc{ref-jiangExpertFeatureengineeringVs2018}{2018}) created
various detection models using different deep learning architectures, as
well as traditional algorithms with expert-engineered features. As with
their evaluation of affect detectors, they found with off-task behavior
detection that the deep models only outperformed the more traditional
algorithms some of the time.

\subsubsection{Other uses}\label{other-uses}

Other direct uses of deep learning for educational data science include
automatic teacher evaluation
(\citeproc{ref-hongmeiApplicationResearchBP2013}{Hongmei, 2013};
\citeproc{ref-rashidLecturerPerformanceSystem2016}{Rashid \& Ahmad,
2016}), optimized spaced repetition algorithms for memory retention
(\citeproc{ref-reddyAcceleratingHumanLearning2017}{Reddy et al., 2017}),
and automatic translation in online chat for collaborative learning
(\citeproc{ref-satoExaminingImpactAutomated2018}{Sato et al., 2018}).
Fernández Alemán et al.
(\citeproc{ref-fernandezalemanEvaluatingStudentResponse2010}{2010}) also
used deep learning to cluster students into one of various `states of
knowledge,' depending on their level of understanding of the subject
matter (computer science, in this case). They then provided targeted
feedback to students based on the results of the clustering, which were
continuously updated as students answered more questions.

\subsection{Indirect uses of deep learning in
education}\label{indirect-uses-of-deep-learning-in-education}

\subsubsection{Feature extraction}\label{feature-extraction}

Deep learning can be used to generate useful features from raw data that
can then be extracted and used as inputs in a different machine learning
model for a separate task. This process, also known as representation
learning, takes advantage of neural networks' ability to pick out
patterns in raw data that correspond to latent features that are helpful
for a prediction task.

For example, L. Jiang \& Bosch
(\citeproc{ref-jiangMiningAssessingAnomalies2022}{2022}) trained a
neural network to detect anomalous student activities, the results of
which were then used as features in a random forest regressor that
predicted course grade. Similarly, Karimi et al.
(\citeproc{ref-karimiOnlineAcademicCourse2020}{2020}) used a GNN to
extract separate student and course embeddings (vectors that map
concepts to a meaningful space) from a knowledge graph of student-course
relations, while using an LSTM to encode students' sequential behavioral
data. They then concatenated these outputs and used them as features in
a classifier that predicted student-course performance. Shi, Shah, et
al. (\citeproc{ref-shiSemiautomaticMisconceptionDiscovery2021}{2021})
used code2vec embeddings
(\citeproc{ref-alonCode2vecLearningDistributed2019}{Alon et al., 2019})
to identify student misconceptions by clustering the embeddings. Azcona
et al. (\citeproc{ref-azconaUser2code2vecEmbeddingsProfiling2019}{2019})
further used these embeddings to represent students based on their code.
To automatically grade students' responses to open-ended math problems,
sentence-BERT---a sentence-level embedding method that keeps semantic
information
(\citeproc{ref-reimersSentenceBERTSentenceEmbeddings2019}{Reimers \&
Gurevych, 2019})---has been used alongside traditional machine learning
models (\citeproc{ref-baralImprovingAutomatedScoring2021}{Baral et al.,
2021}).

\subsubsection{Computer vision}\label{computer-vision}

Another common indirect use of deep learning in education is in computer
vision. Deep learning models have found extraordinary success in this
area over the last few years, far surpassing the accuracy of other
models. In the domain of education, deep-learning-based computer vision
algorithms have been used to detect students' affective states by
leveraging the front camera of a tablet to capture facial cues
(\citeproc{ref-wampflerImageReconstructionTablet2020}{Wampfler et al.,
2020}). Likewise, some research has used OpenPose, a deep learning pose
tracker (\citeproc{ref-caoRealtimeMultiperson2D2017}{Cao et al., 2017}),
to capture motion tracking information of students and teachers
(\citeproc{ref-hurTrackingIndividualsClassroom2022}{Hur \& Bosch,
2022}). This information has in turn been used to create meaningful
features for measuring student collaboration
(\citeproc{ref-hurInformingExpertFeature2023}{Hur et al., 2023}).
Computer vision has even been used to automatically detect what young
students are looking at in classroom video data, making it easier to
measure levels of attentiveness (\citeproc{ref-aungWhoAreThey2018}{Aung
et al., 2018}).

\subsubsection{Automatic speech
recognition}\label{automatic-speech-recognition}

Automatic speech recognition (ASR) has also been used in educational
research. As with computer vision, this is another area where deep
learning is unquestionably the leading technique. This technology has
been used to transcribe noisy classroom speech in collaborative learning
settings, the results of which have been used to analyze group formation
and collaborative problem solving (CPS) skills
(\citeproc{ref-pughSayWhatAutomatic2021}{Pugh et al., 2021};
\citeproc{ref-southwellChallengesFeasibilityAutomatic2022}{Southwell et
al., 2022}; \citeproc{ref-taoAudiobasedGroupDetection2019}{Tao et al.,
2019}). It has also been used in the domain of language learning to
allow learners to interact with their environments more easily in their
target language (\citeproc{ref-nguyenUserorientedEFLSpeaking2018}{T.-H.
Nguyen et al., 2018}). Panaite et al.
(\citeproc{ref-panaiteIdentifyingReadingStrategies2018}{2018})
incorporated ASR into the self-explanation module of ReaderBench
(\citeproc{ref-dascaluReaderBenchEnvironmentAnalyzing2013}{Dascalu et
al., 2013}) so that students can record oral self-explanations of their
understanding of a text and have instant feedback on their
comprehension. Besides transcription, some of these tools have the
ability to capture prosodic features and identify individual speakers
(\citeproc{ref-southwellChallengesFeasibilityAutomatic2022}{Southwell et
al., 2022}).

\subsubsection{Automated qualitative
coding}\label{automated-qualitative-coding}

Using NLP techniques that leverage the power of neural networks to
analyze language, deep learning has also been used to automatically code
student chat logs in a computer-supported collaborative learning (CSCL)
setting (\citeproc{ref-shibataAutomaticCodingCollaborative2017}{Shibata
et al., 2017}). This is typically a time-consuming task that makes it
possible to conduct analyses for a variety of educational research
purposes.

\section{Looking to the future of deep learning in
education}\label{looking-to-the-future-of-deep-learning-in-education}

Extrapolating from the many ways deep learning has been used for
educational data science, and taking into account its advantages and
limitations, we suggest important directions that may allow the field to
make the most of this technology. We describe three primary areas of
focus: further research on model transparency, making contributions to
our understanding of the learning process, and expanding the use of deep
neural networks outside of the research laboratory to impact learners.

\subsection{Increasing trust through greater
transparency}\label{increasing-trust-through-greater-transparency}

The issue of trust is central to ensuring that models are adopted and
used by stakeholders such as parents and teachers. Without trusting that
an algorithm is not only accurate but fair, deep learning models for
education will have a very difficult time breaking out of the lab and
into classrooms and homes. The limitation of diminished interpretability
discussed earlier is a roadblock to increasing trust. This is why there
is increasing interest in creating more transparent models that can be
more easily understood, audited, and learned from.

In the machine learning world, this situation has given rise to
explainable AI (XAI) methods, which are beginning to be explored in
educational data science
(\citeproc{ref-cohauszWhenProbabilitiesAre2022}{Cohausz, 2022}; see Liu
et al.~in this volume; also
\citeproc{ref-somExplainableStudentGroup2021}{Som et al., 2021};
\citeproc{ref-swamyEvaluatingExplainersBlackbox2022}{Swamy et al.,
2022}). While post-hoc explainability methods may not be as transparent
as intrinsically interpretable models
(\citeproc{ref-rudinStopExplainingBlack2019}{Rudin, 2019}), they provide
some transparency to the process by allowing a better understanding of
the inputs, features, and instances that lead to a model's predictions.
This is a helpful step in understanding \emph{why} a model predicts that
a student is likely to behave or learn a certain way
(\citeproc{ref-hurUsingMachineLearning2022}{Hur et al., 2022}). In many
cases, these post-hoc approaches are the only currently existing way to
make deep learning models more transparent due to their intrinsic
complexity. However, the field has yet to converge on a set of
explainability best practices, with fundamental problems of existing
approaches still unaddressed
(\citeproc{ref-swamyFutureHumancentricEXplainable2023}{Swamy et al.,
2023}).

Along with advancing the work of explaining a model's inner workings,
there is the need to develop novel and robust ways to quantify and
evaluate both the interpretability of a model itself and the use of
specific interpretations/explanations in real-world settings. This
requires first understanding the scenarios in which transparency is
important in education, as well as the needs of end users in these
cases. The taxonomy proposed by Doshi-Velez \& Kim
(\citeproc{ref-doshi-velezRigorousScienceInterpretable2017}{2017}) can
be a good starting point, providing a common lexicon for researchers.
From most specific and costly to least robust and resource-intensive,
they organized interpretability evaluation methods into
application-grounded, human-grounded, and functionally grounded
evaluations.

As with other concepts borrowed from disciplines such as machine
learning, educational data science will encounter its own challenges
specific to the use cases, concerns, and stakeholders in this field. One
might imagine a near future in which researchers frequently report not
only a deep learning model's accuracy measures but also its
interpretability evaluations or specific explanations of its
decision-making process. This may lead the field to confidently create
models and explanations that are not merely accurate but also
trustworthy and useful, overcoming the black box problem currently
holding back deep learning for educational data science.

\subsection{Making contributions to learning
theory}\label{making-contributions-to-learning-theory}

As evidenced in the survey section of this chapter, deep learning has
thus far been used in education primarily for practical applications
designed to help with teaching and learning. However, there has been
very little focus on contributing to learning theory. While uses of deep
learning are being informed by our latest understanding of the learning
process, this research has not yet contributed much in return.

It is true that the literature has already identified some interesting
phenomena. From a practical perspective, outcomes related to student
success can often be predicted with surprising accuracy based on just a
few early results (\citeproc{ref-iqbalEarlyStudentGrade2019}{Iqbal et
al., 2019}; \citeproc{ref-murataEarlyDetectionAtrisk2021}{Murata et al.,
2021}). What can this teach us about the learning process or about our
educational systems? The same has been found for predictors of some
learning behaviors, such as wheel spinning
(\citeproc{ref-zhangEarlyDetectionWheel2019}{C. Zhang et al., 2019}).
What this says about the nature of persistence and its relation to
learning has yet to be explored. While these findings have come as a
result of optimizing predictions, there are plenty of reasons to also
consider how they might inform our theories.

Many predictive models have been designed with the hopes of eventually
being used to either directly intervene and guide students or to provide
helpful information to instructors. Some, however, also have the
potential to aid the research process itself. Detection of behaviors and
characteristics such as affect can theoretically make it easier for
researchers to study these areas and their impact on learning. The same
can be said of various aspects of collaborative learning, such as the
dynamics of group formation
(\citeproc{ref-taoAudiobasedGroupDetection2019}{Tao et al., 2019}), the
development of CPS skills (\citeproc{ref-pughSayWhatAutomatic2021}{Pugh
et al., 2021};
\citeproc{ref-southwellChallengesFeasibilityAutomatic2022}{Southwell et
al., 2022}), or the existence of socially shared regulation of learning
behaviors (\citeproc{ref-nguyenExploringSociallyShared2022}{A. Nguyen et
al., 2022}). Likewise, traditional machine learning models have been
used to predict mind wandering among students based on the speech
patterns and behaviors of instructors
(\citeproc{ref-boschQuantifyingClassroomInstructor2018}{Bosch et al.,
2018}; \citeproc{ref-gliserSoundInattentionPredicting2020}{Gliser et
al., 2020}). Chounta et al.
(\citeproc{ref-chountaModelingZoneProximal2017}{2017}) also attempted to
model students' zone of proximal developments (ZPD) using KT, though not
with deep learning. If the deep learning trend continues and its use
expands to replace other algorithms in educational data science, one can
expect that it will begin to make more contributions to our overall
understanding of theories of learning and engagement.

\subsection{Exploring real-world uses beyond the
lab}\label{exploring-real-world-uses-beyond-the-lab}

As with the previous question, actual uses `in the wild' of deep
learning for education are not nearly as ubiquitous as the amount of
research on the topic would have one think. Knowledge tracing is a
perfect example of this. It is possibly the singular use of deep
learning that has attracted the most attention in educational data
science, judging by the number of publications and the need for recent
surveys on the topic (see
\citeproc{ref-abdelrahmanKnowledgeTracingSurvey2023}{Abdelrahman et al.,
2023}; \citeproc{ref-liuSurveyKnowledgeTracing2023}{Q. Liu et al.,
2023}; \citeproc{ref-sarsaEmpiricalEvaluationDeep2022a}{Sarsa et al.,
2022}). Yet the KT algorithms currently being used by intelligent
tutoring systems to track students' knowledge, suggest next problems,
and provide feedback continue to be traditional approaches such as BKT.
This may be partly due to the challenge of interpretability and partly
due to the fast-changing pace of the technology. The widespread
availability of large datasets has made it so that educational data
science researchers can easily try new advanced methods without the need
to collect data or interact even indirectly with students. While this
allows for rapid prototyping without danger of negatively affecting
students, it also means that there has been little opportunity to
actually support learning or teaching.

The most common educational uses of deep learning out in the real world
are the indirect uses of the technology described previously, such as
speech recognition or automatic translation services. The recent
excitement surrounding advanced chatbots such as ChatGPT---a
transformer-based model---has brought a lot of attention to both the
positive and negative potential of such tools to affect how we learn or
to alter the way students are assessed
(\citeproc{ref-baidoo-anuEducationEraGenerative2023}{Baidoo-Anu \& Owusu
Ansah, 2023};
\citeproc{ref-garcia-penalvoPerceptionArtificialIntelligence2023}{García-Peñalvo,
2023}). Similarly, the recent explosion of interest surrounding
generative text-to-image tools, such as DALL-E
(\citeproc{ref-rameshZeroshotTexttoimageGeneration2021}{Ramesh et al.,
2021}) or Stable Diffusion
(\citeproc{ref-rombachHighresolutionImageSynthesis2022}{Rombach et al.,
2022}) may prove to have implications for art educators or aspiring art
students. These are tools that are already widely available to learners.
At present, however, it is impossible to predict how these and
forthcoming deep learning tools may shape the way people obtain and
process new knowledge.

In terms of existing research in education, very little has been
experimental in nature. In a rare, early attempt, Fernández Alemán et
al. (\citeproc{ref-fernandezalemanEvaluatingStudentResponse2010}{2010})
found that their approach to automatic feedback---which used a simple
neural network architecture---successfully helped the students in the
experimental group pass their exam when compared with those in the
control group. T.-H. T.-H. Nguyen et al.
(\citeproc{ref-nguyenUserorientedEFLSpeaking2018}{2018}) also
demonstrated how a mobile app for authentic target-language practice
using automatic speech recognition and feedback can improve students'
motivation and confidence. On the topic of experimental design, Sales et
al. (\citeproc{ref-salesUsingBigData2018}{2018}) creatively used a deep
neural network to improve data usability in randomized control trials in
an effort to bring more causal modeling to this type of educational
research. However, these real-world experiments are still quite rare. On
the other hand, the \emph{proposed}, \emph{hypothetical} scenarios for
supporting learners and instructors with deep learning models are many.

Knowledge tracing algorithms are typically pursued with the intention of
presenting students with the most appropriate problems for their current
levels of understanding and of allowing students and/or instructors to
track learning progress through open learner models and learning
analytics dashboards (\citeproc{ref-bullThereAreOpen2020}{Bull, 2020};
\citeproc{ref-valleStayingTargetSystematic2021}{Valle et al., 2021}).
Many ITSs already perform these tasks to a certain degree. Research into
deep learning knowledge tracing often highlights implications for
implementing the proposed state-of-the-art models for the same purposes
(\citeproc{ref-aiConceptawareDeepKnowledge2019}{Ai et al., 2019};
\citeproc{ref-tanBiDKTDeepKnowledge2022}{W. Tan et al., 2022}).

Likewise, automated assessment is already used in real life learning
scenarios, including high-stakes exams such as the Graduate Record
Examination (GRE;
\citeproc{ref-attaliAutomatedScoringShortanswer2008}{Attali et al.,
2008}), but deep learning is not typically the standard method in
practice. We have described research that shows evidence in favor of
implementing deep learning more widely for this purpose
(\citeproc{ref-taghipourNeuralApproachAutomated2016}{Taghipour \& Ng,
2016}; \citeproc{ref-taySkipFlowIncorporatingNeural2018}{Tay et al.,
2018}; \citeproc{ref-zhaoMemoryaugmentedNeuralModel2017}{Zhao et al.,
2017}), though it remains to be seen if it is eventually adopted.
Automated assessment can help instructors save time grading to have more
time for targeted support, it can help students receive timely feedback,
and it can help tackle problems of scale in large online courses.
Botarleanu et al.
(\citeproc{ref-botarleanuAutomatedSummaryScoring2021}{2021}) have even
suggested that automatic essay scoring could benefit students get an
early idea of the score they would receive on an essay, giving them the
chance to iterate before submission.

Milani et al.
(\citeproc{ref-milaniIntelligentTutoringStrategies2020}{2020}) designed
an intelligent tutoring system that specifically tailors its tutoring
policies to students of different abilities using reinforcement
learning, which can have a direct impact on these populations. They do
this by predicting questions that fall within a student's ZPD, in a
similar vein as the ZPD modeling done by Chounta et al.
(\citeproc{ref-chountaModelingZoneProximal2017}{2017}). Some have also
suggested providing feedback to students and instructors about different
CPS skills being used in collaborative work in order to make possible
more targeted intervention (\citeproc{ref-pughSayWhatAutomatic2021}{Pugh
et al., 2021};
\citeproc{ref-southwellChallengesFeasibilityAutomatic2022}{Southwell et
al., 2022}). Interventions that encourage academically productive talk
(APT) have been shown to enhance learning in CSCL settings
(\citeproc{ref-tegosPromotingAcademicallyProductive2015}{Tegos et al.,
2015}). Other proposed uses of deep learning include informing
instructors of the weak points in their teaching style
(\citeproc{ref-rashidLecturerPerformanceSystem2016}{Rashid \& Ahmad,
2016}), giving students useful ideas about courses to take based on
their goals (\citeproc{ref-jiangGoalbasedCourseRecommendation2019}{W.
Jiang et al., 2019}), optimizing students' learning gains through
personalized spaced repetition algorithms
(\citeproc{ref-reddyAcceleratingHumanLearning2017}{Reddy et al., 2017}),
providing feedback on students' code even when it contains syntax errors
(a difficult task since most code feedback methods rely on abstract
syntax trees;
\citeproc{ref-bhatiaNeurosymbolicProgramCorrector2018}{Bhatia et al.,
2018}), and providing targeted feedback early based on students'
misconceptions
(\citeproc{ref-shiSemiautomaticMisconceptionDiscovery2021}{Shi, Shah, et
al., 2021}).

As with contributions to theory, it remains to be seen if deep learning
methods begin to be more widely implemented in real-world applications
that directly support learners and instructors in coming years.

\section{Conclusion}\label{conclusion}

In this chapter, we discussed a technology that has been at the
forefront of some of the most exciting and powerful uses of artificial
intelligence in recent years---deep artificial neural networks, or deep
learning---and its uses and potential for the field of educational data
science. We provided a brief introduction to deep learning, described
some of its advantages and limitations, presented a survey of its many
uses in education, and discussed how it may further come to shape the
field in the near future. Like any technology, deep learning faces
challenges with implications for its use and adoption, including the
justified mistrust it can create due to its often inscrutable nature.
However, as researchers continue to make headway on creating highly
accurate and interpretable deep neural networks, the potential of deep
learning for educational data science makes the rapid advancements in
this area worth keeping an eye on.

\section*{References}\label{bibliography}
\addcontentsline{toc}{section}{References}

\phantomsection\label{refs}


\begin{CSLReferences}{1}{0}
\bibitem[\citeproctext]{ref-abdelrahmanKnowledgeTracingSequential2019}
Abdelrahman, G., \& Wang, Q. (2019). Knowledge tracing with sequential
key-value memory networks. \emph{Proceedings of the 42nd {International
ACM SIGIR Conference} on {Research} and {Development} in {Information
Retrieval}}, 175--184. \url{https://doi.org/10.1145/3331184.3331195}

\bibitem[\citeproctext]{ref-abdelrahmanKnowledgeTracingSurvey2023}
Abdelrahman, G., Wang, Q., \& Nunes, B. (2023). Knowledge tracing: {A}
survey. \emph{ACM Computing Surveys}, \emph{55}(11), 1--37.
\url{https://doi.org/10.1145/3569576}

\bibitem[\citeproctext]{ref-aiConceptawareDeepKnowledge2019}
Ai, F., Chen, Y., \& Guo, Y. (2019). Concept-aware deep knowledge
tracing and exercise recommendation in an online learning system.
\emph{Proceedings of {The} 12th {International Conference} on
{Educational Data Mining} ({EDM} 2019)}, 240--245.

\bibitem[\citeproctext]{ref-alamSurveyDeepNeural2020}
Alam, M., Samad, M. D., Vidyaratne, L., Glandon, A., \& Iftekharuddin,
K. M. (2020). Survey on deep neural networks in speech and vision
systems. \emph{Neurocomputing}, \emph{417}, 302--321.
\url{https://doi.org/10.1016/j.neucom.2020.07.053}

\bibitem[\citeproctext]{ref-alomStateoftheartSurveyDeep2019}
Alom, M. Z., Taha, T. M., Yakopcic, C., Westberg, S., Sidike, P.,
Nasrin, M. S., Hasan, M., Van Essen, B. C., Awwal, A. A. S., \& Asari,
V. K. (2019). A state-of-the-art survey on deep learning theory and
architectures. \emph{Electronics}, \emph{8}(3), 292.
\url{https://doi.org/10.3390/electronics8030292}

\bibitem[\citeproctext]{ref-alonCode2vecLearningDistributed2019}
Alon, U., Zilberstein, M., Levy, O., \& Yahav, E. (2019). Code2vec:
Learning distributed representations of code. \emph{Proceedings of the
ACM on Programming Languages}, \emph{3}(POPL), 1--29.
\url{https://doi.org/10.1145/3290353}

\bibitem[\citeproctext]{ref-alzubaidiReviewDeepLearning2021}
Alzubaidi, L., Zhang, J., Humaidi, A. J., Al-Dujaili, A., Duan, Y.,
Al-Shamma, O., Santamaría, J., Fadhel, M. A., Al-Amidie, M., \& Farhan,
L. (2021). Review of deep learning: Concepts, {CNN} architectures,
challenges, applications, future directions. \emph{Journal of Big Data},
\emph{8}(1), 53. \url{https://doi.org/10.1186/s40537-021-00444-8}

\bibitem[\citeproctext]{ref-attaliAutomatedScoringShortanswer2008}
Attali, Y., Powers, D., Freedman, M., Harrison, M., \& Obetz, S. (2008).
\emph{Automated scoring of short-answer open-ended {GRE} subject test
items} (\{\{ETS GRE Board Research Report\}\} No. 04-02). Educational
Testing Service,.

\bibitem[\citeproctext]{ref-aungWhoAreThey2018}
Aung, A. M., Ramakrishnan, A., \& Whitehill, J. R. (2018). Who are they
looking at? {Automatic} eye gaze following for classroom observation
video analysis. \emph{Proceedings of the 11th {International Conference}
on {Educational Data Mining}}, 252--258.

\bibitem[\citeproctext]{ref-azconaUser2code2vecEmbeddingsProfiling2019}
Azcona, D., Arora, P., Hsiao, I.-H., \& Smeaton, A. (2019).
User2code2vec: {Embeddings} for profiling students based on
distributional representations of source code. \emph{Proceedings of the
9th {International Conference} on {Learning Analytics} \& {Knowledge}},
86--95. \url{https://doi.org/10.1145/3303772.3303813}

\bibitem[\citeproctext]{ref-baidoo-anuEducationEraGenerative2023}
Baidoo-Anu, D., \& Owusu Ansah, L. (2023). \emph{Education in the era of
generative artificial intelligence ({AI}): Understanding the potential
benefits of {ChatGPT} in promoting teaching and learning}. SSRN
Electronic Journal. \url{https://doi.org/10.2139/ssrn.4337484}

\bibitem[\citeproctext]{ref-bakerMoreAccurateStudent2008}
Baker, R. S. J. d., Corbett, A. T., \& Aleven, V. (2008). More accurate
student modeling through contextual estimation of slip and guess
probabilities in bayesian knowledge tracing. In B. P. Woolf, E. Aïmeur,
R. Nkambou, \& S. Lajoie (Eds.), \emph{Intelligent {Tutoring Systems}}
(pp. 406--415). Springer.
\url{https://doi.org/10.1007/978-3-540-69132-7_44}

\bibitem[\citeproctext]{ref-bakerStateEducationalData2009}
Baker, R. S. J. d, \& Yacef, K. (2009). The state of educational data
mining in 2009: A review and future visions. \emph{JEDM {\textbar}
Journal of Educational Data Mining}, \emph{1}(1), 3--17.
\url{https://doi.org/gg2gfp}

\bibitem[\citeproctext]{ref-bakerLabelingStudentBehavior2008}
Baker, R. S., \& de Carvalho, A. M. J. A. (2008). Labeling student
behavior faster and more precisely with text replays. \emph{Proceedings
of the 1st {International Conference} on {Educational Data Mining}
({EDM})}, 38--47.

\bibitem[\citeproctext]{ref-bakerAlgorithmicBiasEducation2022}
Baker, R. S., \& Hawn, A. (2022). Algorithmic bias in education.
\emph{International Journal of Artificial Intelligence in Education},
\emph{32}(4), 1052--1092.
\url{https://doi.org/10.1007/s40593-021-00285-9}

\bibitem[\citeproctext]{ref-baralImprovingAutomatedScoring2021}
Baral, S., Botelho, A. F., \& Erickson, J. A. (2021). Improving
automated scoring of student open responses in mathematics.
\emph{Proceedings of {The} 14th {International Conference} on
{Educational Data Mining} ({EDM21})}, 130--138.

\bibitem[\citeproctext]{ref-beckWheelspinningStudentsWho2013}
Beck, J. E., \& Gong, Y. (2013). Wheel-spinning: {Students} who fail to
master a skill. In H. C. Lane, K. Yacef, J. Mostow, \& P. Pavlik (Eds.),
\emph{Artificial {Intelligence} in {Education}} (pp. 431--440).
Springer. \url{https://doi.org/10.1007/978-3-642-39112-5_44}

\bibitem[\citeproctext]{ref-bhatiaNeurosymbolicProgramCorrector2018}
Bhatia, S., Kohli, P., \& Singh, R. (2018). Neuro-symbolic program
corrector for introductory programming assignments. \emph{Proceedings of
the 40th {International Conference} on {Software Engineering}}, 60--70.
\url{https://doi.org/10.1145/3180155.3180219}

\bibitem[\citeproctext]{ref-boschQuantifyingClassroomInstructor2018}
Bosch, N., Mills, C., Wammes, J. D., \& Smilek, D. (2018). Quantifying
classroom instructor dynamics with computer vision. In C. Penstein Rosé,
R. Martínez-Maldonado, H. U. Hoppe, R. Luckin, M. Mavrikis, K.
Porayska-Pomsta, B. McLaren, \& B. du Boulay (Eds.), \emph{Artificial
{Intelligence} in {Education}} (Vol. 10947, pp. 30--42). Springer
International Publishing.
\url{https://doi.org/10.1007/978-3-319-93843-1_3}

\bibitem[\citeproctext]{ref-botarleanuAutomatedSummaryScoring2021}
Botarleanu, R.-M., Dascalu, M., Allen, L. K., Crossley, S. A., \&
McNamara, D. S. (2021). Automated summary scoring with {ReaderBench}. In
A. I. Cristea \& C. Troussas (Eds.), \emph{Intelligent {Tutoring
Systems}} (Vol. 12677, pp. 321--332). Springer International Publishing.
\url{https://doi.org/10.1007/978-3-030-80421-3_35}

\bibitem[\citeproctext]{ref-botelhoImprovingSensorfreeAffect2017}
Botelho, A. F., Baker, R. S., \& Heffernan, N. T. (2017). Improving
sensor-free affect detection using deep learning. \emph{Artificial
{Intelligence} in {Education}}, \emph{10331}, 40--51.
\url{https://doi.org/10.1007/978-3-319-61425-0_4}

\bibitem[\citeproctext]{ref-botelhoStudyingAffectDynamics2018}
Botelho, A. F., Baker, R. S., Ocumpaugh, J., \& Heffernan, N. T. (2018).
Studying affect dynamics and chronometry using sensor-free detectors.
\emph{Proceedings of the 11th {International Conference} on {Educational
Data Mining}}, 157--166.

\bibitem[\citeproctext]{ref-botelhoDeepLearningDeep2022}
Botelho, A. F., Prihar, E., \& Heffernan, N. T. (2022). Deep learning or
deep ignorance? {Comparing} untrained recurrent models in educational
contexts. In M. M. Rodrigo, N. Matsuda, A. I. Cristea, \& V. Dimitrova
(Eds.), \emph{Artificial {Intelligence} in {Education}: 23rd
{International Conference}, {AIED} 2022} (Vol. 13355, pp. 281--293).
Springer International Publishing.
\url{https://doi.org/10.1007/978-3-031-11644-5_23}

\bibitem[\citeproctext]{ref-botelhoDevelopingEarlyDetectors2019}
Botelho, A. F., Varatharaj, A., Patikorn, T., Doherty, D., Adjei, S. A.,
\& Beck, J. E. (2019). Developing early detectors of student attrition
and wheel spinning using deep learning. \emph{IEEE Transactions on
Learning Technologies}, \emph{12}(2), 158--170.
\url{https://doi.org/10.1109/TLT.2019.2912162}

\bibitem[\citeproctext]{ref-bullThereAreOpen2020}
Bull, S. (2020). There are open learner models about! \emph{IEEE
Transactions on Learning Technologies}, \emph{13}(2), 425--448.
\url{https://doi.org/10.1109/TLT.2020.2978473}

\bibitem[\citeproctext]{ref-caderPotentialUseDeep2020}
Cader, A. (2020). The potential for the use of deep neural networks in
e-learning student evaluation with new data augmentation method.
\emph{Artificial Intelligence in Education}, \emph{12164}, 37--42.
\url{https://doi.org/10.1007/978-3-030-52240-7_7}

\bibitem[\citeproctext]{ref-caoRealtimeMultiperson2D2017}
Cao, Z., Simon, T., Wei, S.-E., \& Sheikh, Y. (2017). Realtime
multi-person {2D} pose estimation using part affinity fields. \emph{2017
{IEEE Conference} on {Computer Vision} and {Pattern Recognition}
({CVPR})}, 1302--1310. \url{https://doi.org/10.1109/CVPR.2017.143}

\bibitem[\citeproctext]{ref-chenApplyingRecentInnovations2020}
Chen, C., \& Pardos, Z. (2020). Applying recent innovations from {NLP}
to {MOOC} student course trajectory modeling. \emph{Proceedings of {The}
13th {International Conference} on {Educational Data Mining} ({EDM}
2020)}, 581--585.

\bibitem[\citeproctext]{ref-choiAppropriateQueryKey2020}
Choi, Y., Lee, Y., Cho, J., Baek, J., Kim, B., Cha, Y., Shin, D., Bae,
C., \& Heo, J. (2020). Towards an appropriate query, key, and value
computation for knowledge tracing. \emph{Proceedings of the {Seventh ACM
Conference} on {Learning} @ {Scale}}, 341--344.
\url{https://doi.org/10.1145/3386527.3405945}

\bibitem[\citeproctext]{ref-chountaModelingZoneProximal2017}
Chounta, I.-A., McLaren, B. M., Albacete, P., Jordan, P., \& Katz, S.
(2017). Modeling the zone of proximal development with a computational
approach. \emph{Proceedings of the 10th {International Conference} on
{Educational Data Mining} ({EDM} 2017)}, 56--57.

\bibitem[\citeproctext]{ref-cohauszWhenProbabilitiesAre2022}
Cohausz, L. (2022). When probabilities are not enough: A framework for
causal explanations of student success models. \emph{Journal of
Educational Data Mining}, \emph{14}(3), 52--75.
\url{https://doi.org/10.5281/zenodo.7304800}

\bibitem[\citeproctext]{ref-corbettKnowledgeTracingModeling1995}
Corbett, A. T., \& Anderson, J. R. (1995). Knowledge tracing: {Modeling}
the acquisition of procedural knowledge. \emph{User Modelling and
User-Adapted Interaction}, \emph{4}(4), 253--278.
\url{https://doi.org/b4wwjx}

\bibitem[\citeproctext]{ref-craigAffectLearningExploratory2004}
Craig, S., Graesser, A., Sullins, J., \& Gholson, B. (2004). Affect and
learning: {An} exploratory look into the role of affect in learning with
{AutoTutor}. \emph{Journal of Educational Media}, \emph{29}(3),
241--250. \url{https://doi.org/10.1080/1358165042000283101}

\bibitem[\citeproctext]{ref-dmelloDynamicsAffectiveStates2012}
D'Mello, S., \& Graesser, A. (2012). Dynamics of affective states during
complex learning. \emph{Learning and Instruction}, \emph{22}(2),
145--157. \url{https://doi.org/10.1016/j.learninstruc.2011.10.001}

\bibitem[\citeproctext]{ref-dascaluReaderBenchEnvironmentAnalyzing2013}
Dascalu, M., Dessus, P., Trausan-Matu, Ş., Bianco, M., \& Nardy, A.
(2013). {ReaderBench}, an environment for analyzing text complexity and
reading strategies. In D. Hutchison, T. Kanade, J. Kittler, J. M.
Kleinberg, F. Mattern, J. C. Mitchell, M. Naor, O. Nierstrasz, C. Pandu
Rangan, B. Steffen, M. Sudan, D. Terzopoulos, D. Tygar, M. Y. Vardi, G.
Weikum, H. C. Lane, K. Yacef, J. Mostow, \& P. Pavlik (Eds.),
\emph{{AIED} 13 - 16th {In-} ternational {Conference} on {Artificial
Intelligence} in {Education}} (Vol. 7926, pp. 379--388). Springer Berlin
Heidelberg. \url{https://doi.org/10.1007/978-3-642-39112-5_39}

\bibitem[\citeproctext]{ref-dascaluReaderBenchMultilingualFramework2017}
Dascalu, M., Gutu, G., Ruseti, S., Paraschiv, I. C., Dessus, P.,
McNamara, D. S., Crossley, S. A., \& Trausan-Matu, S. (2017).
{ReaderBench}: {A} multi-lingual framework for analyzing text
complexity. In É. Lavoué, H. Drachsler, K. Verbert, J. Broisin, \& M.
Pérez-Sanagustín (Eds.), \emph{Data {Driven Approaches} in {Digital
Education}, {Proc} 12th {European Conference} on {Technology Enhanced
Learning}, {EC-TEL} 2017} (Vol. 10474, pp. 495--499). Springer
International Publishing.
\url{https://doi.org/10.1007/978-3-319-66610-5_48}

\bibitem[\citeproctext]{ref-delianidiStudentPerformancePrediction2021}
Delianidi, M., Diamantaras, K., Chrysogonidis, G., \& Nikiforidis, V.
(2021, June). Student performance prediction using dynamic neural
models. \emph{Proceedings of {The} 14th {International Conference} on
{Educational Data Mining} ({EDM} 2021)}.
\url{https://arxiv.org/abs/2106.00524}

\bibitem[\citeproctext]{ref-dingWhyDeepKnowledge2019}
Ding, X., \& Larson, E. C. (2019). Why deep knowledge tracing has less
depth than anticipated. \emph{Proceedings of {The} 12th {International
Conference} on {Educational Data Mining} ({EDM} 2019)}, 282--287.

\bibitem[\citeproctext]{ref-doleckPredictiveAnalyticsEducation2020}
Doleck, T., Lemay, D. J., Basnet, R. B., \& Bazelais, P. (2020).
Predictive analytics in education: {A} comparison of deep learning
frameworks. \emph{Education and Information Technologies}, \emph{25}(3),
1951--1963. \url{https://doi.org/10.1007/s10639-019-10068-4}

\bibitem[\citeproctext]{ref-doshi-velezRigorousScienceInterpretable2017}
Doshi-Velez, F., \& Kim, B. (2017). \emph{Towards a rigorous science of
interpretable machine learning} (No. arXiv:1702.08608). arXiv.
\url{https://arxiv.org/abs/1702.08608}

\bibitem[\citeproctext]{ref-ericksonAutomatedGradingStudent2020}
Erickson, J. A., Botelho, A. F., McAteer, S., Varatharaj, A., \&
Heffernan, N. T. (2020). The automated grading of student open responses
in mathematics. \emph{Proceedings of the {Tenth International
Conference} on {Learning Analytics} \& {Knowledge}}, 615--624.
\url{https://doi.org/10.1145/3375462.3375523}

\bibitem[\citeproctext]{ref-feiTemporalModelsPredicting2015}
Fei, M., \& Yeung, D.-Y. (2015). Temporal models for predicting student
dropout in massive open online courses. \emph{2015 {IEEE International
Conference} on {Data Mining Workshop} ({ICDMW})}, 256--263.
\url{https://doi.org/10.1109/ICDMW.2015.174}

\bibitem[\citeproctext]{ref-fengUnderstandingDropoutsMOOCs2019}
Feng, W., Tang, J., \& Liu, T. X. (2019). Understanding dropouts in
{MOOCs}. \emph{Proceedings of the AAAI Conference on Artificial
Intelligence}, \emph{33}(01), 517--524.
\url{https://doi.org/10.1609/aaai.v33i01.3301517}

\bibitem[\citeproctext]{ref-fernandezalemanEvaluatingStudentResponse2010}
Fernández Alemán, J. L., Palmer-Brown, D., \& Draganova, C. (2010).
Evaluating student response driven feedback in a programming course.
\emph{2010 10th {IEEE International Conference} on {Advanced Learning
Technologies}}, 279--283. \url{https://doi.org/10.1109/ICALT.2010.82}

\bibitem[\citeproctext]{ref-garcia-penalvoPerceptionArtificialIntelligence2023}
García-Peñalvo, F. J. (2023). The perception of artificial intelligence
in educational contexts after the launch of {ChatGPT}: Disruption or
panic? \emph{Education in the Knowledge Society (EKS)}, \emph{24},
e31279. \url{https://doi.org/10.14201/eks.31279}

\bibitem[\citeproctext]{ref-gervetWhenDeepLearning2020}
Gervet, T., Schneider, J., Koedinger, K., \& Mitchell, T. (2020). When
is deep learning the best approach to knowledge tracing? \emph{Journal
of Educational Data Mining}, \emph{12}(3), 31--54.

\bibitem[\citeproctext]{ref-ghoshContextawareAttentiveKnowledge2020}
Ghosh, A., Heffernan, N., \& Lan, A. S. (2020). Context-aware attentive
knowledge tracing. \emph{Proceedings of the 26th {ACM SIGKDD
International Conference} on {Knowledge Discovery} \& {Data Mining}},
2330--2339. \url{https://doi.org/10.1145/3394486.3403282}

\bibitem[\citeproctext]{ref-gliserSoundInattentionPredicting2020}
Gliser, I., Mills, C., Bosch, N., Smith, S., Smilek, D., \& Wammes, J.
D. (2020). The sound of inattention: Predicting mind wandering with
automatically derived features of instructor speech. \emph{Artificial
{Intelligence} in {Education}}, \emph{12163}, 204--215.
\url{https://doi.org/10.1007/978-3-030-52237-7_17}

\bibitem[\citeproctext]{ref-gonzalez-brenesGeneralFeaturesKnowledge2014}
Gonzalez-Brenes, J., Huang, Y., \& Brusilovsky, P. (2014). General
features in knowledge tracing to model multiple subskills, temporal item
response theory, and expert knowledge. \emph{The 7th {International
Conference} on {Educational Data Mining}}, 84--91.

\bibitem[\citeproctext]{ref-gottfriedChronicAbsenteeismIts2014}
Gottfried, M. A. (2014). Chronic absenteeism and its effects on
students' academic and socioemotional outcomes. \emph{Journal of
Education for Students Placed at Risk (JESPAR)}, \emph{19}(2), 53--75.
\url{https://doi.org/10.1080/10824669.2014.962696}

\bibitem[\citeproctext]{ref-hongmeiApplicationResearchBP2013}
Hongmei, L. (2013). Application research of {BP} neural network in
{English} teaching evaluation. \emph{TELKOMNIKA Indonesian Journal of
Electrical Engineering}, \emph{11}(8), 4602--4608.
\url{https://doi.org/10.11591/telkomnika.v11i8.3085}

\bibitem[\citeproctext]{ref-huAcademicPerformanceEstimation2019}
Hu, Q., \& Rangwala, H. (2019). Academic performance estimation with
attention-based graph convolutional networks. \emph{Proceedings of {The}
12th {International Conference} on {Educational Data Mining} ({EDM}
2019)}, 69--78.

\bibitem[\citeproctext]{ref-huAssessingStudentContributions2020}
Hu, T., Sun, G., \& Xu, Z. (2020). Assessing student contributions in
wiki-based collaborative writing system. \emph{Proceedings of {The} 13th
{International Conference} on {Educational Data Mining} ({EDM} 2020)},
615--619.

\bibitem[\citeproctext]{ref-hurTrackingIndividualsClassroom2022}
Hur, P., \& Bosch, N. (2022). Tracking individuals in classroom videos
via post-processing {OpenPose} data. \emph{{LAK22}: 12th {International
Learning Analytics} and {Knowledge Conference}}, 465--471.
\url{https://doi.org/10.1145/3506860.3506888}

\bibitem[\citeproctext]{ref-hurUsingMachineLearning2022}
Hur, P., Lee, H., Bhat, S., \& Bosch, N. (2022). Using machine learning
explainability methods to personalize interventions for students.
\emph{Proceedings of the 15th {International Conference} on {Educational
Data Mining}}, 438--445. \url{https://doi.org/10.5281/ZENODO.6853181}

\bibitem[\citeproctext]{ref-hurInformingExpertFeature2023}
Hur, P., Machaka, N., Krist, C., \& Bosch, N. (2023). Informing expert
feature engineering through automated approaches: Implications for
coding qualitative classroom video data. \emph{{LAK23}}.
\url{https://doi.org/10.1145/3576050.3576090}

\bibitem[\citeproctext]{ref-iqbalEarlyStudentGrade2019}
Iqbal, Z., Qayyum, A., Latif, S., \& Qadir, J. (2019). Early student
grade prediction: An empirical study. \emph{2019 2nd {International
Conference} on {Advancements} in {Computational Sciences} ({ICACS})},
1--7. \url{https://doi.org/10.23919/ICACS.2019.8689136}

\bibitem[\citeproctext]{ref-jarbouDeepLearningbasedSchool2022}
Jarbou, M., Won, D., Gillis-Mattson, J., \& Romanczyk, R. (2022). Deep
learning-based school attendance prediction for autistic students.
\emph{Scientific Reports}, \emph{12}.
\url{https://doi.org/10.1038/s41598-022-05258-z}

\bibitem[\citeproctext]{ref-jiangMiningAssessingAnomalies2022}
Jiang, L., \& Bosch, N. (2022). Mining and assessing anomalies in
students' online learning activities with self-supervised machine
learning. \emph{Proceedings of the 15th {International Conference} on
{Educational Data Mining}}, 549--554.
\url{https://doi.org/10.5281/ZENODO.6852948}

\bibitem[\citeproctext]{ref-jiangGoalbasedCourseRecommendation2019}
Jiang, W., Pardos, Z. A., \& Wei, Q. (2019). Goal-based course
recommendation. \emph{Proceedings of the 9th {International Conference}
on {Learning Analytics} \& {Knowledge}}, 36--45.
\url{https://doi.org/10.1145/3303772.3303814}

\bibitem[\citeproctext]{ref-jiangExpertFeatureengineeringVs2018}
Jiang, Y., Bosch, N., Baker, R. S., Paquette, L., Ocumpaugh, J., Andres,
J. Ma. A. L., Moore, A. L., \& Biswas, G. (2018). Expert
feature-engineering vs. Deep neural networks: {Which} is better for
sensor-free affect detection? In C. Penstein Rosé, R.
Martínez-Maldonado, H. U. Hoppe, R. Luckin, M. Mavrikis, K.
Porayska-Pomsta, B. McLaren, \& B. du Boulay (Eds.), \emph{Artificial
{Intelligence} in {Education}} (pp. 198--211). Springer International
Publishing. \url{https://doi.org/10.1007/978-3-319-93843-1_15}

\bibitem[\citeproctext]{ref-kanterDeepFeatureSynthesis2015}
Kanter, J. M., \& Veeramachaneni, K. (2015). Deep feature synthesis:
{Towards} automating data science endeavors. \emph{2015 {IEEE
International Conference} on {Data Science} and {Advanced Analytics}
({DSAA})}, 1--10. \url{https://doi.org/10.1109/DSAA.2015.7344858}

\bibitem[\citeproctext]{ref-karimiOnlineAcademicCourse2020}
Karimi, H., Derr, T., Huang, J., \& Tang, J. (2020). Online academic
course performance prediction using relational graph convolutional
neural network. \emph{Proceedings of {The} 13th {International
Conference} on {Educational Data Mining} ({EDM} 2020)}, 444--450.

\bibitem[\citeproctext]{ref-karumbaiahUsingNeuralNetworkbased2022}
Karumbaiah, S., Zhang, J., Baker, R. S., Scruggs, R., Cade, W.,
Clements, M., \& Lin, S. (2022). Using neural network-based knowledge
tracing for a learning system with unreliable skill tags.
\emph{Proceedings of the 15th {International Conference} on {Educational
Data Mining}}, 333--338. \url{https://doi.org/10.5281/ZENODO.6852994}

\bibitem[\citeproctext]{ref-khajahIntegratingLatentfactorKnowledgetracing2014}
Khajah, M. M., Wing, R. M., Lindsey, R. V., \& Mozer, M. C. (2014).
Integrating latent-factor and knowledge-tracing models to predict
individual differences in learning. \emph{In {In} Submission}.

\bibitem[\citeproctext]{ref-khajahHowDeepKnowledge2016}
Khajah, M., Lindsey, R. V., \& Mozer, M. C. (2016). How deep is
knowledge tracing? \emph{Proceedings of the 9th {International
Conference} on {Educational Data Mining}}, 94--101.

\bibitem[\citeproctext]{ref-kimGritNetStudentPerformance2018}
Kim, B.-H., Vizitei, E., \& Ganapathi, V. (2018). {GritNet}: {Student
Performance Prediction} with {Deep Learning}. \emph{Proceedings of the
11th {International Conference} on {Educational Data Mining}}.

\bibitem[\citeproctext]{ref-kizilcecAlgorithmicFairnessEducation2022}
Kizilcec, R. F., \& Lee, H. (2022). Algorithmic fairness in education.
In \emph{The {Ethics} of {Artificial Intelligence} in {Education}}.
Routledge.

\bibitem[\citeproctext]{ref-lanAccurateInterpretableSensorfree2020}
Lan, A. S., Botelho, A., Karumbaiah, S., Baker, R. S., \& Heffernan, N.
(2020). Accurate and interpretable sensor-free affect detectors via
monotonic neural networks. \emph{Companion {Proceedings} 10th
{International Conference} on {Learning Analytics} \& {Knowledge}
({LAK20})}.

\bibitem[\citeproctext]{ref-linComparisonsBKTRNN2017}
Lin, C., \& Chi, M. (2017). A comparisons of {BKT}, {RNN} and {LSTM} for
learning gain prediction. In E. André, R. Baker, X. Hu, Ma. M. T.
Rodrigo, \& B. du Boulay (Eds.), \emph{Artificial {Intelligence} in
{Education}} (Vol. 10331, pp. 536--539). Springer International
Publishing. \url{https://doi.org/10.1007/978-3-319-61425-0_58}

\bibitem[\citeproctext]{ref-liuEKTExerciseawareKnowledge2021}
Liu, Q., Huang, Z., Yin, Y., Chen, E., Xiong, H., Su, Y., \& Hu, G.
(2021). {EKT}: {Exercise-aware} knowledge tracing for student
performance prediction. \emph{IEEE Transactions on Knowledge and Data
Engineering}, \emph{33}(1), 100--115.
\url{https://doi.org/10.1109/TKDE.2019.2924374}

\bibitem[\citeproctext]{ref-liuSurveyKnowledgeTracing2023}
Liu, Q., Shen, S., Huang, Z., Chen, E., \& Zheng, Y. (2023). A survey of
knowledge tracing. \emph{IEEE Transactions on Knowledge and Data
Engineering}. \url{https://arxiv.org/abs/2105.15106}

\bibitem[\citeproctext]{ref-liuImprovingKnowledgeTracing2020}
Liu, Y., Yang, Y., Chen, X., Shen, J., Zhang, H., \& Yu, Y. (2020).
Improving knowledge tracing via pre-training question embeddings.
\emph{Proceedings of the {Twenty-Ninth International Joint Conference}
on {Artificial Intelligence}}, 1577--1583.
\url{https://doi.org/10.24963/ijcai.2020/219}

\bibitem[\citeproctext]{ref-livierisPredictingStudentsPerformance2012}
Livieris, I. E., Drakopoulou, K., \& Pintelas, P. (2012). Predicting
students' performance using artificial neural networks.
\emph{Proceedings of Information and Communication Technologies in
Education}, 321--328.

\bibitem[\citeproctext]{ref-livierisPredictingSecondarySchool2019}
Livieris, I. E., Drakopoulou, K., Tampakas, V. T., Mikropoulos, T. A.,
\& Pintelas, P. (2019). Predicting secondary school students'
performance utilizing a semi-supervised learning approach. \emph{Journal
of Educational Computing Research}, \emph{57}(2), 448--470.
\url{https://doi.org/10.1177/0735633117752614}

\bibitem[\citeproctext]{ref-longAutomaticalGraphbasedKnowledge2022}
Long, T., Liu, Y., Zhang, W., Xia, W., He, Z., Tang, R., \& Yu, Y.
(2022). Automatical graph-based knowledge tracing. In A. Mitrovic \& N.
Bosch (Eds.), \emph{Proceedings of the 15th {International Conference}
on {Educational Data Mining}} (pp. 710--714). Zenodo.
\url{https://doi.org/10.5281/ZENODO.6853057}

\bibitem[\citeproctext]{ref-maoDeepLearningVs2018}
Mao, Y., Chi, M., \& Lin, C. (2018). Deep learning vs. Bayesian
knowledge tracing: Student models for interventions. \emph{Journal of
Educational Data Mining}, \emph{10}(2), 28--54.

\bibitem[\citeproctext]{ref-matsudaHowQuicklyCan2016}
Matsuda, N., Chandrasekaran, S., \& Stamper, J. (2016). How quickly can
wheel spinning be detected? \emph{Proceedings of {The} 9th
{International Conference} on {Educational Data Mining} ({EDM} 2016)},
607--608.

\bibitem[\citeproctext]{ref-milaniIntelligentTutoringStrategies2020}
Milani, S., Fan, Z., Gulati, S., Nguyen, T., Yadav, A., \& Fang, F.
(2020). Intelligent tutoring strategies for students with autism
spectrum disorder: {A} reinforcement learning approach.
\emph{{AI4EDU-20}: {Workshop} on {Artifical Intelligence} for
{Education} Held at the 34nd {AAAI Conference} on {Artificial
Intelligence} ({AAAI})}, 7.

\bibitem[\citeproctext]{ref-munkhaugenIndividualCharacteristicsStudents2019}
Munkhaugen, E. K., Torske, T., Gjevik, E., Nærland, T., Pripp, A. H., \&
Diseth, T. H. (2019). Individual characteristics of students with autism
spectrum disorders and school refusal behavior. \emph{Autism},
\emph{23}(2), 413--423. \url{https://doi.org/10.1177/1362361317748619}

\bibitem[\citeproctext]{ref-murataEarlyDetectionAtrisk2021}
Murata, R., Shimada, A., \& Minematsu, T. (2021). Early detection of
at-risk students based on knowledge distillation {RNN} models.
\emph{Proceedings of {The} 14th {International Conference} on
{Educational Data Mining} ({EDM21})}.

\bibitem[\citeproctext]{ref-nakagawaGraphbasedKnowledgeTracing2019}
Nakagawa, H., Iwasawa, Y., \& Matsuo, Y. (2019). Graph-based knowledge
tracing: {Modeling} student proficiency using graph neural network.
\emph{{IEEE}/{WIC}/{ACM International Conference} on {Web
Intelligence}}, 156--163. \url{https://doi.org/10.1145/3350546.3352513}

\bibitem[\citeproctext]{ref-namanlooImprovingPeerAssessment2022}
Namanloo, A. A., Thorpe, J., \& Salehi-Abari, A. (2022). Improving peer
assessment with graph neural networks. \emph{Proceedings of the 15th
{International Conference} on {Educational Data Mining}}, 325--332.
\url{https://doi.org/10.5281/ZENODO.6853014}

\bibitem[\citeproctext]{ref-nguyenExploringSociallyShared2022}
Nguyen, A., Järvelä, S., Wang, Y., \& Rosé, C. (2022). Exploring
socially shared regulation with an {AI} deep learning approach using
multimodal data. \emph{Proceedings of {International Conferences} of
{Learning Sciences} ({ICLS})}.

\bibitem[\citeproctext]{ref-nguyenUserorientedEFLSpeaking2018}
Nguyen, T.-H., Hwang, W.-Y., Pham, X.-L., \& Ma, Z.-H. (2018).
User-oriented {EFL} speaking through application and exercise: Instant
speech translation and shadowing in authentic context. \emph{Educational
Technology \& Society}, \emph{21}(4), 129--142.

\bibitem[\citeproctext]{ref-oladokunPredictingStudentsAcademic2008}
Oladokun, V. O. (2008). Predicting students' academic performance using
artificial neural network: {A} case study of an engineering course.
\emph{The Pacific Journal of Science and Technology}, \emph{9}(1),
72--79.

\bibitem[\citeproctext]{ref-panaiteIdentifyingReadingStrategies2018}
Panaite, M., Dascalu, M., Dessus, P., Bianco, M., \& Trausan-Matu, S.
(2018). Identifying reading strategies employed by learners within their
oral {French} self-explanations. \emph{Proc. 5th {Int}. {Workshop} on
``{Semantic} and {Collaborative Technologies} for the {Web}``, Joint to
the 14th {Int}. {Sci}. {Conf}. On {eLearning} and {Software} for
{Education} ({eLSE} 2018)}, 362--368.

\bibitem[\citeproctext]{ref-pandeySelfattentiveModelKnowledge2019}
Pandey, S., \& Karypis, G. (2019). A self-attentive model for knowledge
tracing. \emph{Proceedings of The 12th International Conference on
Educational Data Mining (EDM 2019)}, 284--389.
\url{https://doi.org/10.48550/arXiv.1907.06837}

\bibitem[\citeproctext]{ref-pandeyRKTRelationawareSelfattention2020}
Pandey, S., \& Srivastava, J. (2020). {RKT}: {Relation-aware}
self-attention for knowledge tracing. \emph{Proceedings of the 29th {ACM
International Conference} on {Information} \& {Knowledge Management}},
1205--1214. \url{https://doi.org/10.1145/3340531.3411994}

\bibitem[\citeproctext]{ref-pardosAffectiveStatesState2014}
Pardos, Z. A., Baker, R. S. J. D., San Pedro, M., Gowda, S. M., \&
Gowda, S. M. (2014). Affective states and state tests: {Investigating}
how affect and engagement during the school year predict end-of-year
learning outcomes. \emph{Journal of Learning Analytics}, \emph{1}(1),
107--128. \url{https://doi.org/10.18608/jla.2014.11.6}

\bibitem[\citeproctext]{ref-pardosDeterminingSignificanceItem2009}
Pardos, Z. A., \& Heffernan, N. T. (2009, July). Determining the
significance of item order in randomized problem sets.
\emph{International {Working Group} on {Educational Data Mining}}.

\bibitem[\citeproctext]{ref-pardosModelingIndividualizationBayesian2010}
Pardos, Z. A., \& Heffernan, N. T. (2010). Modeling {Individualization}
in a {Bayesian Networks Implementation} of {Knowledge Tracing}. In P. De
Bra, A. Kobsa, \& D. Chin (Eds.), \emph{User {Modeling}, {Adaptation},
and {Personalization}} (pp. 255--266). Springer.
\url{https://doi.org/b8jr9n}

\bibitem[\citeproctext]{ref-pardosKTIDEMIntroducingItem2011}
Pardos, Z. A., \& Heffernan, N. T. (2011). {KT-IDEM}: {Introducing} item
difficulty to the knowledge tracing model. In J. A. Konstan, R. Conejo,
J. L. Marzo, \& N. Oliver (Eds.), \emph{User {Modeling}, {Adaption} and
{Personalization}} (pp. 243--254). Springer.
\url{https://doi.org/10.1007/978-3-642-22362-4_21}

\bibitem[\citeproctext]{ref-pavlikjrPerformanceFactorsAnalysis2009}
Pavlik Jr, P. I., Cen, H., \& Koedinger, K. R. (2009). Performance
factors analysis -- a new alternative to knowledge tracing.
\emph{Proceedings of the 14th {International Conference} on {Artificial
Intelligence} in {Education}}.

\bibitem[\citeproctext]{ref-penmetsaInvestigateEffectivenessCode2021}
Penmetsa, P., Shi, Y., \& Price, T. (2021). \emph{Investigate
effectiveness of code features in knowledge tracing task on novice
programming course}.

\bibitem[\citeproctext]{ref-piechDeepKnowledgeTracing2015}
Piech, C., Bassen, J., Huang, J., Ganguli, S., Sahami, M., Guibas, L.
J., \& Sohl-Dickstein, J. (2015). Deep knowledge tracing. \emph{Advances
in {Neural Information Processing Systems}}, \emph{28}.

\bibitem[\citeproctext]{ref-pintoInterpretableNeuralNetworks2023}
Pinto, J. D., Paquette, L., \& Bosch, N. (2023). Interpretable neural
networks vs. Expert-defined models for learner behavior detection.
\emph{Companion {Proceedings} of the 13th {International Conference} on
{Learning Analytics} \& {Knowledge Conference} ({LAK23})}, 105--107.

\bibitem[\citeproctext]{ref-puSelfattentionKnowledgeTracing2022}
Pu, S., \& Becker, L. (2022). Self-attention in knowledge tracing: {Why}
it works. In M. M. Rodrigo, N. Matsuda, A. I. Cristea, \& V. Dimitrova
(Eds.), \emph{Artificial {Intelligence} in {Education}} (Vol. 13355, pp.
731--736). Springer International Publishing.
\url{https://doi.org/10.1007/978-3-031-11644-5_75}

\bibitem[\citeproctext]{ref-pughSayWhatAutomatic2021}
Pugh, S. L., Subburaj, S. K., Rao, A. R., Stewart, A. E. B.,
Andrews-Todd, J., \& D'Mello, S. K. (2021). Say what? {Automatic}
modeling of collaborative problem solving skills from student speech in
the wild. \emph{Proceedings of {The} 14th {International Conference} on
{Educational Data Mining} ({EDM} 2021)}, 55--67.

\bibitem[\citeproctext]{ref-qiuBetterGradePrediction2022}
Qiu, W., Supraja, S., \& Khong, A. W. H. (2022). Toward better grade
prediction via {A2GP}: {An} academic achievement inspired predictive
model. \emph{Proceedings of the 15th {International Conference} on
{Educational Data Mining}}, 195--205.
\url{https://doi.org/10.5281/ZENODO.6852984}

\bibitem[\citeproctext]{ref-rameshZeroshotTexttoimageGeneration2021}
Ramesh, A., Pavlov, M., Goh, G., Gray, S., Voss, C., Radford, A., Chen,
M., \& Sutskever, I. (2021). \emph{Zero-shot text-to-image generation}
(No. arXiv:2102.12092). arXiv.
\url{https://doi.org/10.48550/arXiv.2102.12092}

\bibitem[\citeproctext]{ref-raschProbabilisticModelsIntelligence1993}
Rasch, G. (1993). \emph{Probabilistic models for some intelligence and
attainment tests}. MESA Press.

\bibitem[\citeproctext]{ref-rashidLecturerPerformanceSystem2016}
Rashid, T. A., \& Ahmad, H. A. (2016). Lecturer performance system using
neural network with particle swarm optimization. \emph{Computer
Applications in Engineering Education}, \emph{24}(4), 629--638.
\url{https://doi.org/10.1002/cae.21737}

\bibitem[\citeproctext]{ref-reddyAcceleratingHumanLearning2017}
Reddy, S., Levine, S., \& Dragan, A. (2017). Accelerating human learning
with deep reinforcement learning. \emph{{NeurIPS Workshop} on {Machine
Teaching} 2017}.

\bibitem[\citeproctext]{ref-reimersSentenceBERTSentenceEmbeddings2019}
Reimers, N., \& Gurevych, I. (2019). Sentence-{BERT}: Sentence
embeddings using siamese {BERT-networks}. \emph{Proceedings of the 2019
{Conference} on {Empirical Methods} in {Natural Language Processing} and
the 9th {International Joint Conference} on {Natural Language
Processing} ({EMNLP-IJCNLP})}, 3982--3992.
\url{https://doi.org/10.18653/v1/D19-1410}

\bibitem[\citeproctext]{ref-riordanInvestigatingNeuralArchitectures2017}
Riordan, B., Horbach, A., Cahill, A., Zesch, T., \& Lee, C. M. (2017).
Investigating neural architectures for short answer scoring.
\emph{Proceedings of the 12th {Workshop} on {Innovative Use} of {NLP}
for {Building Educational Applications}}, 159--168.
\url{https://doi.org/10.18653/v1/W17-5017}

\bibitem[\citeproctext]{ref-rodrigoAffectiveBehavioralPredictors2009}
Rodrigo, M. M. T., Baker, R. S., Jadud, M. C., \& Amarra, A. C. M.
(2009). Affective and behavioral predictors of novice programmer
achievement. \emph{{ITiCSE}'09}.

\bibitem[\citeproctext]{ref-rombachHighresolutionImageSynthesis2022}
Rombach, R., Blattmann, A., Lorenz, D., Esser, P., \& Ommer, B. (2022).
High-resolution image synthesis with latent diffusion models. \emph{2022
{IEEE}/{CVF Conference} on {Computer Vision} and {Pattern Recognition}
({CVPR})}, 10674--10685.
\url{https://doi.org/10.1109/CVPR52688.2022.01042}

\bibitem[\citeproctext]{ref-rudinStopExplainingBlack2019}
Rudin, C. (2019). Stop explaining black box machine learning models for
high stakes decisions and use interpretable models instead. \emph{Nature
Machine Intelligence}, \emph{1}(5), 206--215.
\url{https://doi.org/10.1038/s42256-019-0048-x}

\bibitem[\citeproctext]{ref-salesUsingBigData2018}
Sales, A. C., Botelho, A., \& Patikorn, T. (2018). Using big data to
sharpen design-based inference in {A}/{B} tests. \emph{Proceedings of
the 11th {International Conference} on {Educational Data Mining}}.

\bibitem[\citeproctext]{ref-sanpedroPredictingCollegeEnrollment2013}
San Pedro, M. O. Z., Baker, R. S., Bowers, A. J., \& Heffernan, N. T.
(2013). Predicting college enrollment from student interaction with an
intelligent tutoring system in middle school. \emph{Educational {Data
Mining} 2013}.

\bibitem[\citeproctext]{ref-sarsaEmpiricalEvaluationDeep2022a}
Sarsa, S., Leinonen, J., \& Hellas, A. (2022). Empirical evaluation of
deep learning models for knowledge tracing: {Of} hyperparameters and
metrics on performance and replicability. \emph{Journal of Educational
Data Mining}, \emph{14}(2), 32--102.
\url{https://doi.org/10.5281/zenodo.7086179}

\bibitem[\citeproctext]{ref-satoExaminingImpactAutomated2018}
Sato, T., Ogura, M., Aota, S., \& Burden, T. (2018). Examining the
impact of an automated translation chatbot on online collaborative
dialog for incidental {L2} learning. In P. Taalas, J. Jalkanen, L.
Bradley, \& S. Thouësny (Eds.), \emph{Future-proof {CALL}: Language
learning as exploration and encounters -- short papers from {EUROCALL}
2018} (pp. 284--289). Research-publishing.net.
\url{https://doi.org/10.14705/rpnet.2018.26.851}

\bibitem[\citeproctext]{ref-shenAutomaticRecommendationTechnology2016}
Shen, X., Yi, B., Zhang, Z., Shu, J., \& Liu, H. (2016). Automatic
recommendation technology for learning resources with convolutional
neural network. \emph{2016 {International Symposium} on {Educational
Technology} ({ISET})}, 30--34.
\url{https://doi.org/10.1109/ISET.2016.12}

\bibitem[\citeproctext]{ref-shiCodeDKTCodebasedKnowledge2022}
Shi, Y., Chi, M., Barnes, T., \& Price, T. (2022, July). Code-{DKT}: {A}
code-based knowledge tracing model for programming tasks.
\emph{Proceedings of the 15th {International Conference} on {Educational
Data Mining}}. \url{https://doi.org/10.5281/ZENODO.6853105}

\bibitem[\citeproctext]{ref-shiMoreLessExploring2021}
Shi, Y., Mao, Y., Barnes, T., Chi, M., \& Price, T. W. (2021). More with
less: {Exploring} how to use deep learning effectively through
semi-supervised learning for automatic bug detection in student code.
\emph{Proceedings of {The} 14th {International Conference} on
{Educational Data Mining} ({EDM21})}, 446--453.

\bibitem[\citeproctext]{ref-shiSemiautomaticMisconceptionDiscovery2021}
Shi, Y., Shah, K., Wang, W., Marwan, S., Penmetsa, P., \& Price, T.
(2021). Toward semi-automatic misconception discovery using code
embeddings. \emph{{LAK21}: 11th {International Learning Analytics} and
{Knowledge Conference}}, 606--612.
\url{https://doi.org/10.1145/3448139.3448205}

\bibitem[\citeproctext]{ref-shibataAutomaticCodingCollaborative2017}
Shibata, C., Inaba, T., \& Ando, K. (2017). Towards automatic coding of
collaborative learning data with deep learning technology. \emph{{eLmL}
2017:{The Ninth International Conference} on {Mobile}, {Hybrid}, and
{On-line Learning}}, 65--71.

\bibitem[\citeproctext]{ref-shinSAINTIntegratingTemporal2021}
Shin, D., Shim, Y., Yu, H., Lee, S., Kim, B., \& Choi, Y. (2021).
{SAINT}+: {Integrating} temporal features for {EdNet} correctness
prediction. \emph{{LAK21}: 11th {International Learning Analytics} and
{Knowledge Conference}}, 490--496.
\url{https://doi.org/10.1145/3448139.3448188}

\bibitem[\citeproctext]{ref-somExplainableStudentGroup2021}
Som, A., Kim, S., \& Lopez-Prado, B. (2021). Towards explainable student
group collaboration assessment models using temporal representations of
individual student roles. \emph{Proceedings of The 14th International
Conference on Educational Data Mining (EDM21)}, 750--754.

\bibitem[\citeproctext]{ref-songJKTJointGraph2021}
Song, X., Li, J., Tang, Y., Zhao, T., Chen, Y., \& Guan, Z. (2021).
{JKT}: {A} joint graph convolutional network based {Deep Knowledge
Tracing}. \emph{Information Sciences}, \emph{580}, 510--523.
\url{https://doi.org/10.1016/j.ins.2021.08.100}

\bibitem[\citeproctext]{ref-southwellChallengesFeasibilityAutomatic2022}
Southwell, R., Pugh, S., Perkoff, E. M., Clevenger, C., Bush, J.,
Lieber, R., Ward, W., Foltz, P., \& D'Mello, S. (2022). Challenges and
feasibility of automatic speech recognition for modeling student
collaborative discourse in classrooms. \emph{Proceedings of the 15th
{International Conference} on {Educational Data Mining}}, 302--315.
\url{https://doi.org/10.5281/zenodo.6853109}

\bibitem[\citeproctext]{ref-suExerciseenhancedSequentialModeling2018}
Su, Y., Liu, Q., Liu, Q., Huang, Z., Yin, Y., Chen, E., Ding, C., Wei,
S., \& Hu, G. (2018). Exercise-enhanced sequential modeling for student
performance prediction. \emph{Proceedings of the AAAI Conference on
Artificial Intelligence}, \emph{32}(1), 2435--2443.
\url{https://doi.org/10.1609/aaai.v32i1.11864}

\bibitem[\citeproctext]{ref-swamyFutureHumancentricEXplainable2023}
Swamy, V., Frej, J., \& Käser, T. (2023). \emph{The future of
human-centric {eXplainable Artificial Intelligence} ({XAI}) is not
post-hoc explanations} (No. arXiv:2307.00364). arXiv.
\url{https://arxiv.org/abs/2307.00364}

\bibitem[\citeproctext]{ref-swamyEvaluatingExplainersBlackbox2022}
Swamy, V., Radmehr, B., Krco, N., Marras, M., \& Käser, T. (2022).
Evaluating the explainers: Black-box explainable machine learning for
student success prediction in {MOOCs}. \emph{Proceedings of the 15th
{International Conference} on {Educational Data Mining} ({EDM} 2022)}.
\url{https://doi.org/10.5281/ZENODO.6852964}

\bibitem[\citeproctext]{ref-taghipourNeuralApproachAutomated2016}
Taghipour, K., \& Ng, H. T. (2016). A neural approach to automated essay
scoring. \emph{Proceedings of the 2016 {Conference} on {Empirical
Methods} in {Natural Language Processing}}, 1882--1891.
\url{https://doi.org/10.18653/v1/D16-1193}

\bibitem[\citeproctext]{ref-tanAutomaticShortAnswer2020}
Tan, H., Wang, C., Duan, Q., Lu, Y., Zhang, H., \& Li, R. (2020).
Automatic short answer grading by encoding student responses via a graph
convolutional network. \emph{Interactive Learning Environments}, 1--15.
\url{https://doi.org/10.1080/10494820.2020.1855207}

\bibitem[\citeproctext]{ref-tanBiDKTDeepKnowledge2022}
Tan, W., Jin, Y., Liu, M., \& Zhang, H. (2022). {BiDKT}: {Deep}
knowledge tracing with {BERT}. In W. Bao, X. Yuan, L. Gao, T. H. Luan,
\& D. B. J. Choi (Eds.), \emph{Ad {Hoc Networks} and {Tools} for {IT}}
(pp. 260--278). Springer International Publishing.
\url{https://doi.org/10.1007/978-3-030-98005-4_19}

\bibitem[\citeproctext]{ref-tangDeepNeuralNetworks2016}
Tang, S., Peterson, J. C., \& Pardos, Z. A. (2016). Deep neural networks
and how they apply to sequential education data. \emph{Proceedings of
the {Third} (2016) {ACM Conference} on {Learning} @ {Scale}}, 321--324.
\url{https://doi.org/10.1145/2876034.2893444}

\bibitem[\citeproctext]{ref-taoAudiobasedGroupDetection2019}
Tao, Y., Mitsven, S. G., Perry, L. K., Messinger, D. S., \& Shyu, M.-L.
(2019). Audio-based group detection for classroom dynamics analysis.
\emph{2019 {International Conference} on {Data Mining Workshops}
({ICDMW})}, 855--862. \url{https://doi.org/10.1109/ICDMW.2019.00125}

\bibitem[\citeproctext]{ref-tappertWhoFatherDeep2019}
Tappert, C. C. (2019). Who is the father of deep learning? \emph{2019
{International Conference} on {Computational Science} and {Computational
Intelligence} ({CSCI})}, 343--348.
\url{https://doi.org/10.1109/CSCI49370.2019.00067}

\bibitem[\citeproctext]{ref-tatoDeepKnowledgeTracing2022}
Tato, A., \& Nkambou, R. (2022). Deep knowledge tracing on~skills
with~small datasets. In S. Crossley \& E. Popescu (Eds.),
\emph{Intelligent {Tutoring Systems}} (pp. 123--135). Springer
International Publishing.
\url{https://doi.org/10.1007/978-3-031-09680-8_12}

\bibitem[\citeproctext]{ref-taySkipFlowIncorporatingNeural2018}
Tay, Y., Phan, M., Tuan, L. A., \& Hui, S. C. (2018). {SkipFlow}:
{Incorporating} neural coherence features for end-to-end automatic text
scoring. \emph{Proceedings of the AAAI Conference on Artificial
Intelligence}, \emph{32}(1).
\url{https://doi.org/10.1609/aaai.v32i1.12045}

\bibitem[\citeproctext]{ref-tegosPromotingAcademicallyProductive2015}
Tegos, S., Demetriadis, S., \& Karakostas, A. (2015). Promoting
academically productive talk with conversational agent interventions in
collaborative learning settings. \emph{Computers \& Education},
\emph{87}, 309--325. \url{https://doi.org/10.1016/j.compedu.2015.07.014}

\bibitem[\citeproctext]{ref-tongStructurebasedKnowledgeTracing2020}
Tong, S., Liu, Q., Huang, W., Hunag, Z., Chen, E., Liu, C., Ma, H., \&
Wang, S. (2020). Structure-based knowledge tracing: {An} influence
propagation view. \emph{2020 {IEEE International Conference} on {Data
Mining} ({ICDM})}, 541--550.
\url{https://doi.org/10.1109/ICDM50108.2020.00063}

\bibitem[\citeproctext]{ref-valleStayingTargetSystematic2021}
Valle, N., Antonenko, P., Dawson, K., \& Huggins-Manley, A. C. (2021).
Staying on target: A systematic literature review on learner-facing
learning analytics dashboards. \emph{British Journal of Educational
Technology}, \emph{52}(4), 1724--1748.
\url{https://doi.org/10.1111/bjet.13089}

\bibitem[\citeproctext]{ref-vaswaniAttentionAllYou2017}
Vaswani, A., Shazeer, N., Parmar, N., Uszkoreit, J., Jones, L., Gomez,
A. N., Kaiser, Ł., \& Polosukhin, I. (2017). Attention is all you need.
\emph{Advances in {Neural Information Processing Systems}}, \emph{30}.

\bibitem[\citeproctext]{ref-wampflerImageReconstructionTablet2020}
Wampfler, R., Emch, A., Solenthaler, B., \& Gross, M. (2020). Image
reconstruction of tablet front camera recordings in educational
settings. \emph{Proceedings of {The} 13th {International Conference} on
{Educational Data Mining} ({EDM} 2020)}, 245--256.

\bibitem[\citeproctext]{ref-wangLearningRepresentStudent2017}
Wang, L., Sy, A., Liu, L., \& Piech, C. (2017). Learning to represent
student knowledge on programming exercises using deep learning.
\emph{Proceedings of the 10th {International Conference} on {Educational
Data Mining}}, 324--329.

\bibitem[\citeproctext]{ref-wangTimeSeriesClassification2017}
Wang, Z., Yan, W., \& Oates, T. (2017). Time series classification from
scratch with deep neural networks: A strong baseline. \emph{2017
{International Joint Conference} on {Neural Networks} ({IJCNN})},
1578--1585. \url{https://doi.org/10.1109/IJCNN.2017.7966039}

\bibitem[\citeproctext]{ref-whitehillDelvingDeeperMOOC2017}
Whitehill, J., Mohan, K., Seaton, D., Rosen, Y., \& Tingley, D. (2017).
\emph{Delving deeper into {MOOC} student dropout prediction} (No.
arXiv:1702.06404). arXiv. \url{https://arxiv.org/abs/1702.06404}

\bibitem[\citeproctext]{ref-xiaoDetectingProblemStatements2020}
Xiao, Y., Zingle, G., Jia, Q., Shah, H. R., Zhang, Y., Li, T.,
Karovaliya, M., Zhao, W., Song, Y., Ji, J., Balasubramaniam, A., Patel,
H., Bhalasubbramanian, P., Patel, V., \& Gehringer, E. F. (2020).
Detecting problem statements in peer assessments. \emph{Proceedings of
{The} 13th {International Conference} on {Educational Data Mining}
({EDM} 2020)}, 704--709. \url{https://arxiv.org/abs/2006.04532}

\bibitem[\citeproctext]{ref-xiongGoingDeeperDeep2016}
Xiong, X., Zhao, S., Inwegen, V., \& Beck, J. E. (2016). Going deeper
with deep knowledge tracing. \emph{Proceedings of the 9th {International
Conference} on {Educational Data Mining}}, 545--550.

\bibitem[\citeproctext]{ref-yeungIncorporatingFeaturesLearned2019}
Yeung, C.-K., \& Yeung, D.-Y. (2019). Incorporating features learned by
an enhanced deep knowledge tracing model for {STEM}/non-{STEM} job
prediction. \emph{International Journal of Artificial Intelligence in
Education}, \emph{29}(3), 317--341.
\url{https://doi.org/10.1007/s40593-019-00175-1}

\bibitem[\citeproctext]{ref-yudelsonBetterDataBeat2014}
Yudelson, M. V., Fancsali, S. E., Ritter, S., Berman, S. R., Nixon, T.,
\& Joshi, A. (2014). Better {Data Beat Big Data}. \emph{Proceedings of
the 7th {International Conference} on {Educational Data Mining}}.

\bibitem[\citeproctext]{ref-yudelsonIndividualizedBayesianKnowledge2013}
Yudelson, M. V., Koedinger, K. R., \& Gordon, G. J. (2013).
Individualized {Bayesian} knowledge tracing models. In H. C. Lane, K.
Yacef, J. Mostow, \& P. Pavlik (Eds.), \emph{Artificial {Intelligence}
in {Education}} (pp. 171--180). Springer.
\url{https://doi.org/10.1007/978-3-642-39112-5_18}

\bibitem[\citeproctext]{ref-zhangEarlyDetectionWheel2019}
Zhang, C., Huang, Y., Wang, J., Lu, D., Fang, W., Fancsali, S.,
Holstein, K., \& Aleven, V. (2019). Early detection of wheel spinning:
{Comparison} across tutors, models, features, and operationalizations.
\emph{Proceedings of The 12th International Conference on Educational
Data Mining (EDM 2019)}, 468--473.

\bibitem[\citeproctext]{ref-zhangDynamicKeyvalueMemory2017}
Zhang, J., Shi, X., King, I., \& Yeung, D.-Y. (2017). Dynamic key-value
memory networks for knowledge tracing. \emph{Proceedings of the 26th
{International Conference} on {World Wide Web}}, 765--774.
\url{https://doi.org/10.1145/3038912.3052580}

\bibitem[\citeproctext]{ref-zhangNovelNeuralSource2019}
Zhang, J., Wang, X., Zhang, H., Sun, H., Wang, K., \& Liu, X. (2019). A
novel neural source code representation based on abstract syntax tree.
\emph{2019 {IEEE}/{ACM} 41st {International Conference} on {Software
Engineering} ({ICSE})}, 783--794.
\url{https://doi.org/10.1109/ICSE.2019.00086}

\bibitem[\citeproctext]{ref-zhangUndergraduateGradePrediction2021}
Zhang, Y., An, R., Cui, J., \& Shang, X. (2021). Undergraduate grade
prediction in chinese higher education using convolutional neural
networks. \emph{{LAK21}: 11th {International Learning Analytics} and
{Knowledge Conference}}, 462--468.
\url{https://doi.org/10.1145/3448139.3448184}

\bibitem[\citeproctext]{ref-zhaoMemoryaugmentedNeuralModel2017}
Zhao, S., Zhang, Y., Xiong, X., Botelho, A., \& Heffernan, N. (2017). A
memory-augmented neural model for automated grading. \emph{Proceedings
of the {Fourth} (2017) {ACM Conference} on {Learning} @ {Scale}},
189--192. \url{https://doi.org/10.1145/3051457.3053982}

\bibitem[\citeproctext]{ref-zhuangComprehensiveSurveyTransfer2021}
Zhuang, F., Qi, Z., Duan, K., Xi, D., Zhu, Y., Zhu, H., Xiong, H., \&
He, Q. (2021). A comprehensive survey on transfer learning.
\emph{Proceedings of the IEEE}, \emph{109}(1), 43--76.
\url{https://doi.org/10.1109/JPROC.2020.3004555}

\end{CSLReferences}
\end{document}